\documentclass[prb,aps,twocolumn,10pt,showkeys]{revtex4-1}
\usepackage{graphics,graphicx,epsfig,color,latexsym,float,amsmath}
\usepackage[colorlinks,citecolor=blue,linkcolor=blue,urlcolor=blue]{hyperref}
\usepackage[version=4]{mhchem}
	
\newcommand{\abs}[1]{\left| #1 \right|} % for absolute value
\newcommand{\avg}[1]{\left< #1 \right>} % for average
\newcommand{\ket}[1]{\left| #1 \right>} % for Dirac bras
\newcommand{\bra}[1]{\left< #1 \right|} % for Dirac kets
\newcommand{\diag}{\mbox{diag}} % for diag

\begin{document}
\title{Complex magnetic ordering and associated topological Hall effect \\in a two-dimensional metallic chiral magnet}
\author{Nyayabanta Swain, Munir Shahzad and Pinaki Sengupta}
\affiliation{School of Physical and Mathematical Sciences, 
Nanyang Technological University, 21 Nanyang Link, Singapore 637371}
	
\date{\today}

\begin{abstract}
Motivated by recent experiments on the observation of room temperature 
skyrmions in a layered heterostructure and subsequent demonstration of
topological Hall effect in the same system, we have studied a minimal
model of itinerant electrons coupled to local moments with competing
interactions in an external magnetic field. Working in the limit of 
strong magneto-electric coupling where the fast dynamics of the electrons 
can be decoupled from the slow dynamics of the local moments (treated
as classical spins), we analyze the multiple field induced magnetic
phases and the associated electronic transport properties in these
regimes. Our results help understand the microscopic origin of 
the observed phenomena and further provide crucial insight into
unconventional magneto-transport in metallic chiral magnets. 
\end{abstract}

\maketitle

\section{Introduction}\label{sec:intro}
	
The observation of skyrmion phase in the helical magnet \ce{MnSi} has
led to a rapid growth in the study of non-collinear magnetic phases
and associated magneto-electric phenomena~\cite{muhlbauer_skyrmion_2009}. 
First postulated as a topologically protected solution to a class of
non-linear sigma model in high energy physics\cite{Skyrme_HEP}, 
skyrmions arise in magnetic systems as spontaneously formed spin textures 
with non-trivial topology~\cite{Bogdanov_theory_analytical_2001,rosler_spontaneous_2006,nagaosa_topological_2013}.
These spin textures are topologically protected against any deformations~\cite{nagaosa_topological_2013,hagemeister_stability_2015} 
which make them attractive for practical applications~\cite{Wiesendanger_skyrmion_application_2016,Fert_skyrmion_application_2017}. 
In magnetic switching devices, efficient coupling of electric currents 
to these spin textures lead to large spin-transfer torques 
that arise at very low current densities~\cite{jonietz_spin-torque_2010,fert_spin-torque_2013}.
These spin textures drive unusual phenomena such as 
magneto-electric effect for non-collinear magnetic orderings~\cite{wang_strong_2015,gobel_magnetoelectric_2019}
and topological Hall effect for non-coplanar arrangement 
of the spins~\cite{Neubauer_THE_MnSi_2009,kanazawa_MnGe_2011}. 
These spin textures are promising in magnetic memory devices, 
where memory bits can be packed denser and are more robust 
due to their topological nature~\cite{fert1990,romming_skm_control_2013,zhang_skm-logic-gate_2015}. 
Understanding the microscopic origin of these complex structures and 
their associated electronic properties is key for harnessing their unique
functionalities for practical applications~\cite{Fert_skyrmion_application_2017}. 

Skyrmions arise in chiral magnetic systems from a competition
between collinear magnetic order (driven by Heisenberg
exchange and anisotropic interactions) and non-collinearity
caused by Dzyaloshinskii-Moriya (DM) interactions.
Theoretical studies, including  Monte Carlo simulations, 
of the two-dimensional classical spin model demonstrated that 
in chiral magnets with strong DM interaction, skyrmions appear 
in the presence of a magnetic field~\cite{Bogdanov_theory_analytical_1994,Bogdanov_theory_analytical_2001,Nagaosa_theory_MC_2009,Nagaosa_theory_analytical_2010,theory_MC_Ambrose_2013,theory_continum_2016,theory_MC_DFT_Bottcher_2018,theory_simulation_Krauth_2019,theory_simulation_Dagotto_2019}.
Following \ce{MnSi}, skyrmions have been observed in bulk samples 
of other non-centrosymmetric $B20$ compounds~\cite{Wernick_B20-compounds_1972} 
such as \ce{Fe_{1-x}Co_{x}Si}~\cite{munzer_skyrmion_2010,yu_real-space_2010}, 
\ce{FeGe}~\cite{yu_FeGe_2011,wilhelm_FeGe_2011}, 
and \ce{MnGe}~\cite{kanazawa_MnGe_2011,shibata_MnGe_2013,tanigaki_MnGe_2015} 
where intrinsic DM interaction drives non-collinear order. 
Skyrmions also have been observed in an insulating chiral-lattice magnet 
\ce{Cu_2OSeO_3}~\cite{Seki_expt_Cu2OSeO3_2012}.
More recently, artificial heterostructures comprising of thin films
of ferromagnetic materials and heavy metals with strong spin orbit
coupling (SOC) have emerged as a versatile platform
for controllably generating skyrmions\cite{theory_MC_DFT_Dupe2014,theory_MC_DFT_Bottcher_2018}. 
Here broken inversion symmetry
coupled with strong SOC generates an effective DM interaction at the interface
which can be tuned over a wide range by careful material selection. 
Experiments on interfaces of Fe and Fe/Pd thin films 
on Ir(111) system have shown that atomic size skyrmions 
can be engineered with the help of higher-order interactions such as 
the ring exchange~\cite{heinze_spontaneous_2011}
or due to the presence of longer-range competing interactions~\cite{romming_skm_control_2013}.
Furthermore, innovative multi-layer stackings of Ir/Fe/Co/Pt thin films
lead to the strengthening of interfacial DM interactions and 
resulting in small sized skyrmions as compared to the bulk chiral magnets
~\cite{soumyanarayanan_skm_multi-layer_2017,raju_skm_multi-layer_2019}. 
The effect of skyrmions on physical properties is most pronounced 
when their sizes are small. Hence we have used artificially large values of
DM interaction that generates small, lattice scale skyrmions. 
Even though most experimental systems yield skyrmions 
that are much larger than the lattice constants -- we believe 
similar effects will be observed in systems with large sized skyrmions, 
but in weaker forms.

Moving charge carriers in skyrmion hosting materials result in 
novel electrodynamics~\cite{schulz_skm_emergent_electrodynamics_2012}, 
and give rise to the topological Hall effect
(THE)~\cite{Neubauer_THE_MnSi_2009,kanazawa_MnGe_2011,Lee_THE_expt_2009,Li_THE_expt_2013,theory_TSHE_2015,theory_THE_SHE_Manchon_2017}. 
In recent experiments~\cite{soumyanarayanan_skm_multi-layer_2017,raju_skm_multi-layer_2019}, 
the emergence of THE accompanying the onset of the skyrmion phase 
is demonstrated in an artificial heterostructure.

THE is understood in the adiabatic limit to arise from Berry phase accumulated
by electrons moving through a complex non-coplanar spin texture 
with strong magneto-electric coupling~\cite{Nagaosa_SkX_THE_2015,Mertig_SkX_THE_2017,theory_THE_SHE_Denisov2017,gobel_magnetoelectric_2019}. 
As an electron moves through a skyrmion background,
with its spin constrained to be aligned with the local moment at each point, 
it picks up a Berry phase~\cite{Nagaosa_SkX_THE_2015,Mertig_SkX_THE_2017,gobel_family_2018}. 
The gauge flux acts as a fictitious magnetic field acting on the electrons 
and maps the interaction of the electron {\it spin} and local moments 
to a magnetic field coupled to the {\it charge} of the itinerant electrons,
analogous to quantum Hall systems on a lattice. 
The resulting  Lorentz force that drives a transverse current which 
produces distinctive signatures in Hall conductivity that
being promising for potential applications, can be used to
detect the presence of skyrmions~\cite{schulz_skm_emergent_electrodynamics_2012,gobel_magnetoelectric_2019}.

In this paper, we study a metallic chiral magnet with itinerant electrons coupled to
local moments. Working in the limit of strong magneto-electric coupling, the
dynamics of the itinerant electrons and that of the local moments are decoupled.
The magnetic properties are modeled by classical Heisenberg spins on a 
square lattice with ferromagnetic exchange and DM interactions. The ground state
magnetic phases under an external magnetic field are calculated 
using large scale Monte Carlo simulations based on the Metropolis algorithm.  
We focus on various non-collinear magnetic phases stabilized on 
this lattice and their effect on the conduction electron motion 
via a coupled electron-spin model. Our results can be summarized
as follows: $(i)$~Competition between the ferromagnetic exchange and a
strong DM interaction results in a spiral ground state at zero magnetic field.
When an external magnetic field is applied, there is a transition
to a skyrmion crystal (SkX) phase at a critical field, 
and eventually evolves to a fully polarized (FM) phase 
as the applied field strength is increased continuously.
With increasing temperature, the SkX ground state melts to 
a skyrmion liquid (SkL) phase and eventually becomes a paramagnet at high temperature.
$(ii)$~The charge and spin topological Hall conductivities show quantized features in the SkX regime. 
In the ground state, topological Hall conductivity is finite only in the SkX phase,
displaying a discontinuous transition at the spiral-SkX and SkX-fully polarized phase boundaries. 
However, at intermediate temperature, with increasing values of the applied field, 
the topological Hall conductivity varies continuously -- rising from inside the spiral ground state, 
reaches a maximum in the skyrmion phase, and then decaying gradually as the phase changes 
to a fully polarized phase.
These observations are in qualitative agreement with the experiments presented
in Refs.~\onlinecite{soumyanarayanan_skm_multi-layer_2017,raju_skm_multi-layer_2019}, 
providing a plausible microscopic understanding of the
underlying physical mechanism.
Crucially, our results demonstrate that thermodynamic and transport properties
in the the skyrmionic phase (ground state or at finite temperatures) exhibit
distinctive features that can complement the use of more complex imaging
experiments to detect skyrmions.

The rest of the paper is organized as follows. 
In section \ref{sec:model} we discuss the models used in this study.
In section \ref{sec:methods} we describe the method and the observables
we calculate. In section \ref{sec:results} we discuss the detailed results
of our work, followed by the summary in section \ref{sec:summary}.

\section{Model}\label{sec:model}
	
We investigate the ferromagnetic Heisenberg model with additional 
DM interaction and external magnetic field on the square lattice. 
The Hamiltonian under consideration is given by,
	
\begin{equation}
\label{equ:ham-class}
{\mathcal{\hat{H}}_c} = 
-J \sum_{\avg{i,j}} {\bf S}_i \cdot {\bf S}_j 
- \sum_{\avg{i,j}} {\bf D}_{ij} \cdot ({\bf S}_i \times {\bf S}_j) 
-B \sum_i S_i^z
\end{equation}
where $\avg{i,j}$ denotes nearest neighbors and
$J$ is the ferromagnetic exchange coupling. 
${\bf D}_{ij} = D({\bf \hat{z}} \times {\bf \hat{r}}_{ij})$,
with $D$ as the strength of the DM interaction and 
$\mathbf{\hat{r}}_{ij}$ is the unit vector connecting ${\bf r}_i$ to ${\bf r}_j$.
$B$ represents the Zeeman coupling of the spin 
with the external magnetic field applied along the $z-$axis.
The localized spins are treated as classical vectors 
with unit length ($\abs{\mathbf{S}_i}= 1$).
	
Further, to study the properties of conduction electrons 
on the background of localized spin textures, 
we consider the Kondo lattice model given by,
\begin{equation}
\label{equ:ham-elec}
{\mathcal{\hat{H}}_e} = 
-\sum_{\avg{i,j},\sigma} (t_{ij}c_{i\sigma}^\dagger c_{j\sigma}+\mbox{H.c.})
-\mu \sum_i n_i
-J_K \sum_i {\bf S}_i \cdot {\bf s}_i
\end{equation}
where the first term is the kinetic energy, 
involving nearest neighbor hopping $t_{ij}$ and
$c_{i\sigma}^\dagger$ ($c_{i\sigma}$)
is a creation (annihilation) operator 
of an itinerant electron at site ${\bf r}_i$ with spin $\sigma$.
$\mu$ in the second term is the chemical potential 
controlling the density of electrons.
The final term is the Kondo interaction, which describes 
the coupling between the local moment ${\bf S}_i$ 
to the electron spin operator ${\bf s}_i = 
\sum_{\alpha \beta} c^{\dagger}_{i \alpha} {\vec \sigma}_{\alpha \beta} c_{i \beta}$ 
at each site $i$.

\section{Method and observables}\label{sec:methods}
	
We study Hamiltonian (\ref{equ:ham-class}) with classical Monte Carlo (MC)
based on the Metropolis algorithm. We use periodic boundary condition 
on the square lattice of size $L \times L$ with $L=32 - 96$ 
and lattice size increasing in steps of $\Delta L = 8$. 
Efficient thermalization is ensured by a simulated annealing procedure,
where the MC simulation is started from a random spin configuration 
corresponding to high temperature ($T_{high} \sim 2J$), 
and then the temperature is reduced in steps of 
$\Delta T =0.02J$, and $0.01J$ respectively
to reach the lowest temperature of $T_{low} = 0.001J$, 
equilibriating the system at each temperature.
The equilibrium state at $T_{low}$ is used as the ground state 
for calculating physical observable.
At each temperature, we use $5 \times 10^5$ MC sweeps for equilibriation,
and another $5 \times 10^5$ MC sweeps (in steps of 5000 sweeps) 
for calculating the observable in the ground state.
Near the phase boundary, we start with the variational ground state, 
and heat the system using the Monte Carlo method to study its finite temperature aspect. 
As a result the meta-stable phases have been avoided in this regime.

Using the above approach, we explore the magnetic phase
diagram of our model.
The real space configurations of the localized spins 
are used to identify the different magnetic orderings. 
Using the equilibrium MC configurations, 
we compute physical quantities such as magnetization,
static spin structure factor, scalar spin chirality {\it etc.} 
to characterize the magnetic phases and intervening phase transitions.

We define magnetization as,
$$
M=\frac{1}{N} \langle |\sum_{i} {\bf S}_i| \rangle.
$$
The static spin structure factor is the Fourier transform of the 
equal-time spin-spin correlation, given by,
$$
S({\bf Q})=\frac{1}{N}\sum_{i,j}\avg {{\bf S}_i \cdot {\bf S}_j} 
\exp [i{\bf Q} \cdot ({\bf r}_{i} - {\bf r}_{j})]
$$
The skyrmion number is used to identify the skyrmion phase, and
can be inferred from the spin-chirality values, defined as,
$$
\chi = \frac{1}{8\pi N} \langle \sum_{i}
[{\bf S}_i \cdot ({\bf S}_{i+\hat{x}} \times {\bf S}_{i+\hat{y}}) 
+ {\bf S}_i \cdot ({\bf S}_{i-\hat{x}} \times {\bf S}_{i-\hat{y}})] \rangle
$$

% -------------------------------------------------------------------
\begin{figure}[t]
\centering
\includegraphics[width=8.0cm,height=8.0cm]{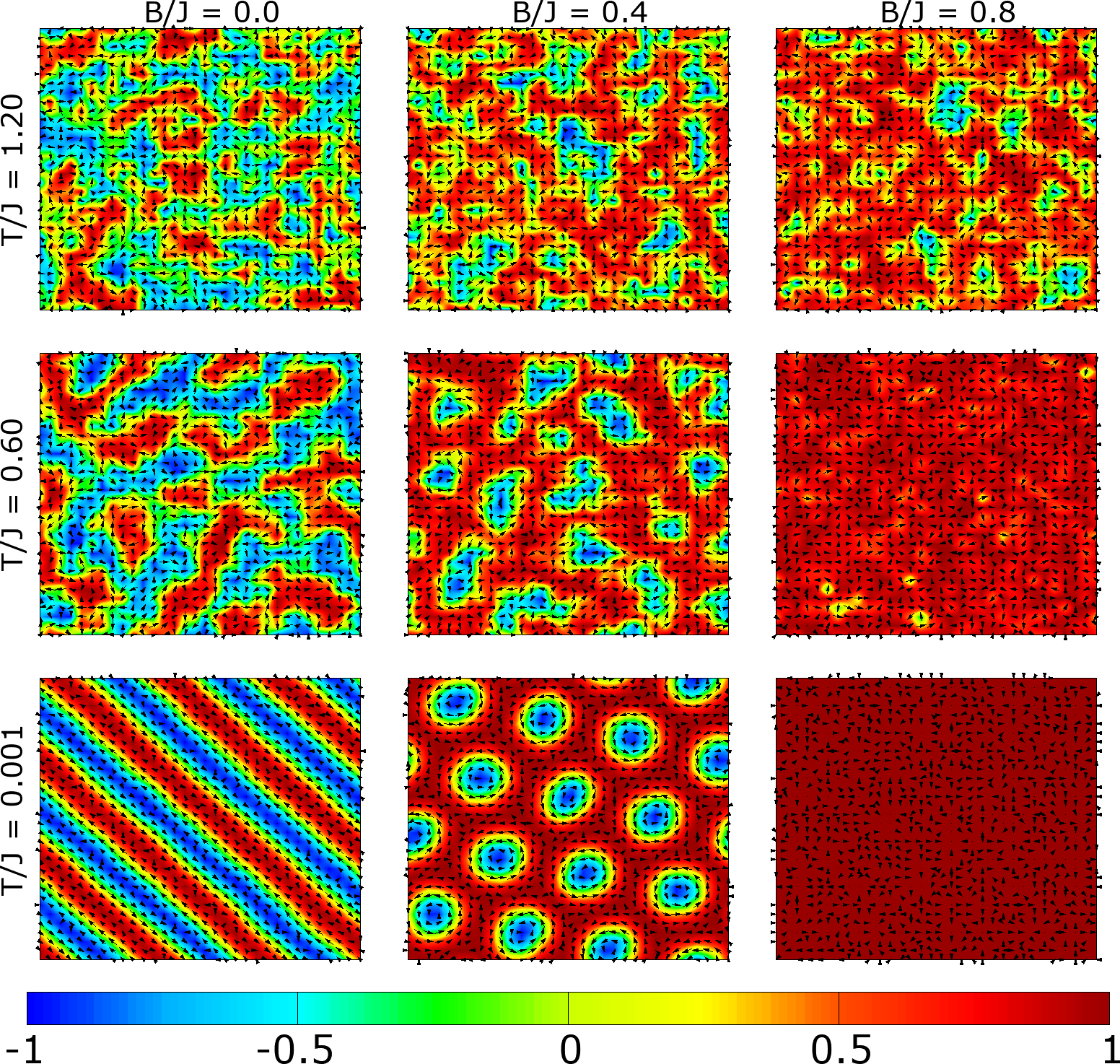}
\caption{\label{config1}
Color online: Representative spin configurations ($ \{ {\bf S}_i \}$) 
obtained for different Zeeman coupling and temperature values 
from our Monte-Carlo simulation for $D = J$.
The $xy$-components are represented by arrows in the $xy$ plane, 
whereas the $z$-component is represented by the color scale.
}
\end{figure}
% -------------------------------------------------------------------

In order to study the transport behavior, we consider
Hamiltonian (\ref{equ:ham-elec}), and calculate 
the transverse conductivity using the Kubo formula~\cite{nagaosa_kubo-formula_AHE_2006},
$$
\sigma_{xy}=\frac{ie^2\hbar}{N}\sum_{\sigma}\sum_{m,n\neq m}(f_m-f_n) \frac{\bra{m}v_x\ket{n}\bra{n}v_y\ket{m}}{(\mathcal{E}_m-\mathcal{E}_n)^2+\eta^2}
$$

\noindent where indices $m$ and $n$ represent the sum over all energy levels, 
$N$ is the total number of sites, 
$f_{m(n)}$ is the Fermi-Dirac distribution function 
for energy $\mathcal{E}_{m(n)}$, 
$\ket{m}$ and $\ket{n}$ are single-particle eigenstates 
corresponding to energy $\mathcal{E}_m$ and 
$\mathcal{E}_n$ and $\eta$ is the scattering rate 
of conduction electrons from the localized spins. 
$v_x$ and $v_y$ are the velocity operators in $\hat{x}$ and $\hat{y}$ directions 
and can be expressed as,
$$
v_\mu=\frac{i}{\hbar}\sum_{j,\sigma}(t_{j,j+\hat{\mu}} c_{j,\sigma}^\dagger c_{j+\hat{\mu},\sigma}- \mbox{H.c.}), \quad \mu=x,y
$$

In addition to the total charge conductivity, we have also calculated the 
contributions from individual spin species.  The non-coplanar spin texture
exerts opposite forces on up and down spin electrons. As we shall discuss
later, this is compounded by the overlap between the two spin species
of electrons in the weak coupling regime, where the spin Hall conductivity
behaves independently of the charge Hall conductivity. The transverse spin
conductivity is given by an analogous Kubo formula that involves the spin current as,
$$
\sigma_{xy}^{S}= \frac{ie}{4\pi N} \sum_{\sigma}\sum_{m,n\neq m}(f_m-f_n) \frac{\bra{m}J_x\ket{n}\bra{n}v_y\ket{m}}{(\mathcal{E}_m-\mathcal{E}_n)^2+\eta^2}
$$
\noindent where $J_x = \frac{1}{2} \{v_x, \diag({\bf S}_1\cdot {\bf \sigma}, ... , {\bf S}_N\cdot {\bf \sigma}) \}$ 
is the spin current operator.

\section{Results}\label{sec:results}
	
\subsection{Magnetic properties}	

{\bf Local spin configuration:} 
Fig.\ref{config1} shows the local
spin configurations for different magnetic field and temperature values 
at fixed $D/J = 1.0$. Since we are working with classical spins, a snapshot
of the local spin configuration from the simulations (after equilibrium) 
provides a visual insight into the nature of the magnetic ground state.
Our simulations are performed with classical spins of unit magnitude 
(i.e $|{\bf S}_i | =1$) at each lattice point.  
We represent the localized spins as follows $-$ 
$S_{ix}$ and $S_{iy}$ are represented by arrows in the $xy$ plane, 
whereas $S_{iz}$ is represented by a color scale (see figure \ref{config1}).
As there is no in-plane anisotropy in our model, 
the configurations obtained as the ground states from the Monte Carlo simulation 
have a companion configuration 
(obtained by rotating all the spins by $\pi/2$ along the $z$ axis)  
with degenerate energy. 

\begin{figure}[t]
\centerline{
\includegraphics[width=7.0cm,height=3.5cm]{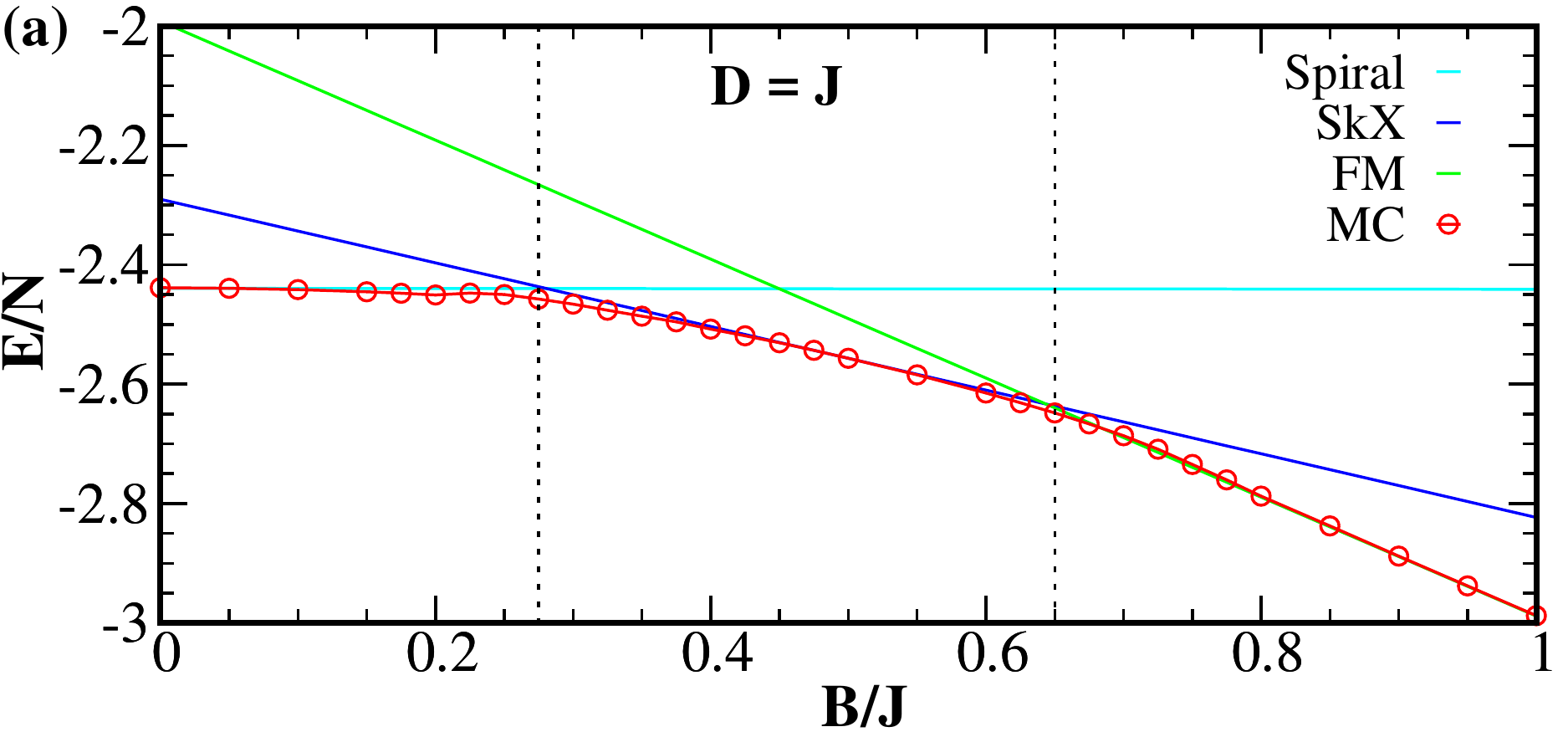}
}
\centerline{
\includegraphics[width=3.5cm,height=2.5cm]{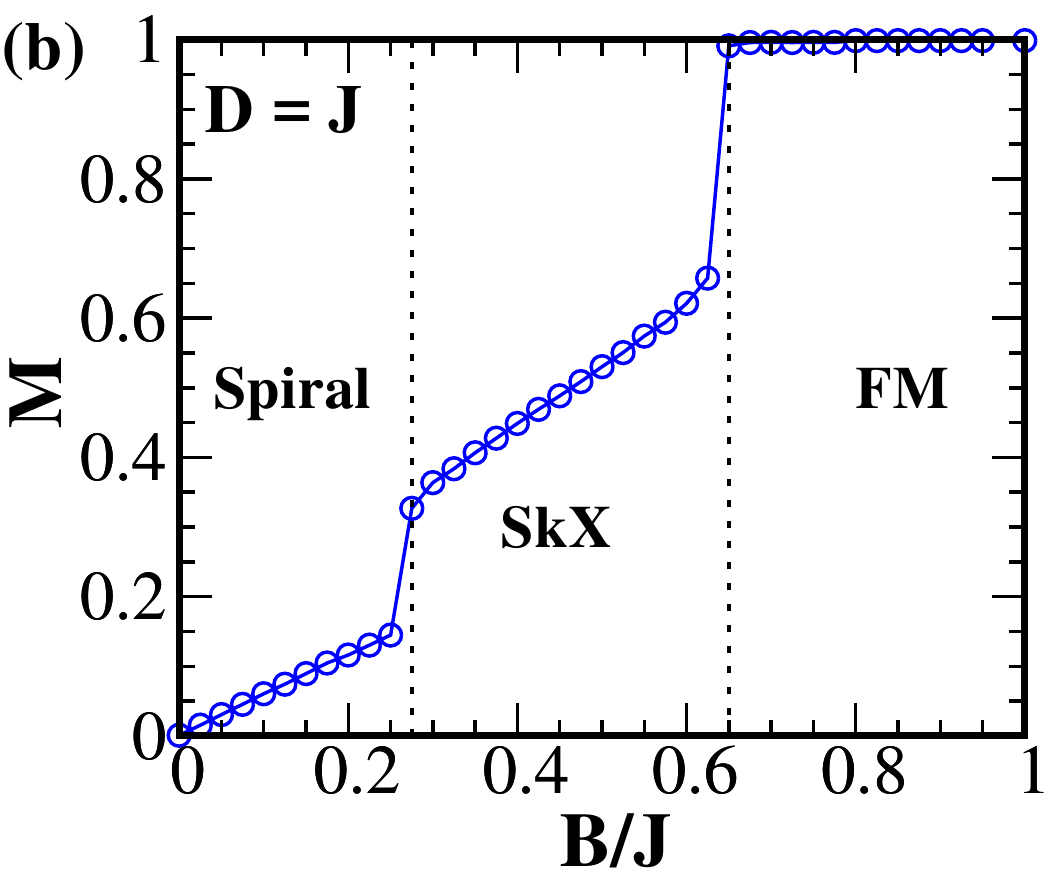}
\includegraphics[width=3.5cm,height=2.5cm]{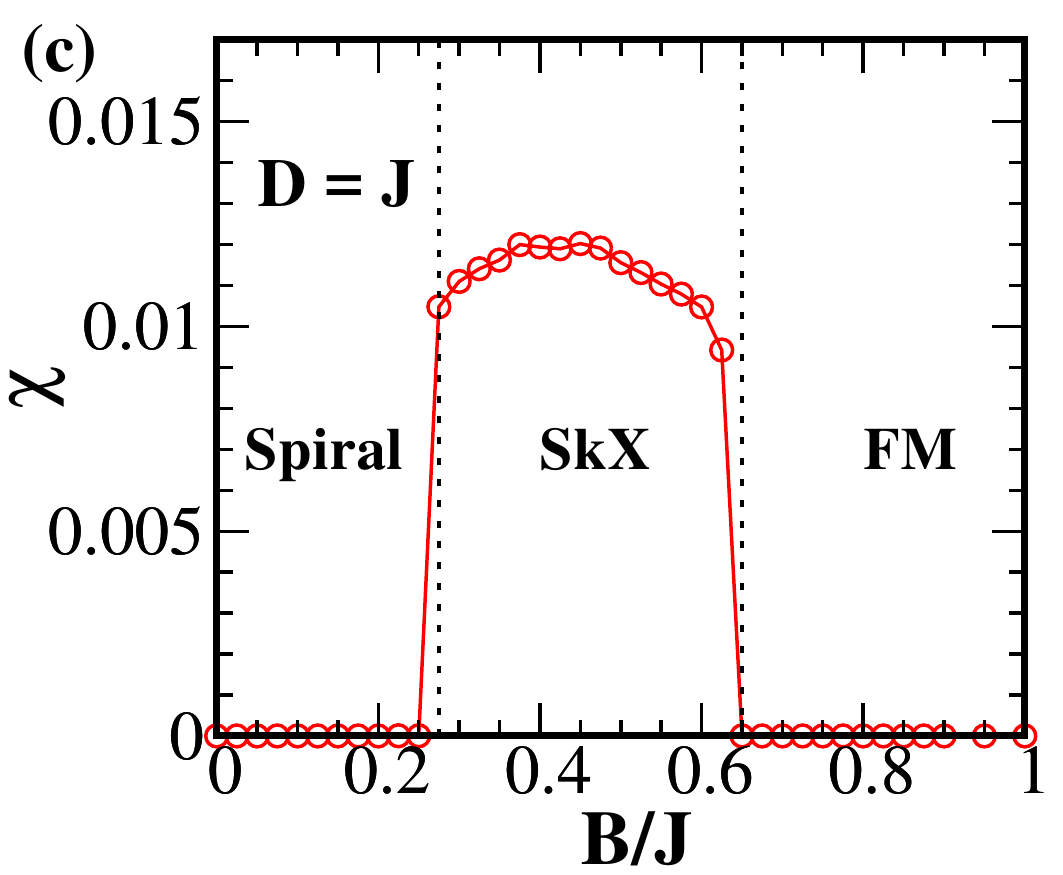}
}
\caption{\label{gs}
Color online: (a)~Energy per lattice site for different phases with changing Zeeman energy.
This is compared with the energy of the configuration
obtained from the Monte Carlo simulation 
at the lowest temperature.
The ground state phases are determined by comparing the
energies of the three static phases and the critical fields  are identified as the points 
of energy level-crossing of the phases.
The spiral phase at $B/J=0.0$, the skyrmion crystal phase 
at $B/J=0.5$ and a fully polarized phase are used as the reference states for carrying out the variational calculation. Comparison with the Monte Carlo data shows that the simulation results are consistent with the variational approach, providing
an important benchmark for the numerical approach used in the study.
(b)~Magnetization calculated from the Monte-Carlo configurations in the ground state
for different Zeeman field values. $M$ increases monotonically with the magnetic field in the spiral and Skyrmion crystal phases and reached full saturation in the 
ferromagnetic phase, changing discontinuously at the phase boundaries. 
(c)~Spin-chirality obtained from MC simulation for the ground state
with varying applied field. $\chi = 0$ both in the spiral and the polarized phases,
and $\chi > 0$ in the skyrmion crystal phase. The chirality also exhibits discontinuous
jumps at the phase boundaries, consistent with the magnetization behavior.
}
\end{figure}

Fig.\ref{gs}(a) shows the energy of the spiral, skyrmion crystal, 
and ferromagnetic phases over the studied Zeeman coupling regime. 
Variational principle suggests that the ground state changes from 
a spiral phase, to a skyrmion crystal, and eventually to a 
field polarized ferromagnetic phase with increasing the Zeeman coupling.
Further, we see the energy of the magnetic phase (corresponding to 
the lowest simulated temperature) obtained from our Monte Carlo simulation 
follows very closely the energy of the variationally obtained ground state phases;
thereby confirming the correctness of the simulation process 
to reach the ground state. The critical magnetic field values 
and the phase transitions in the ground state
are inferred from the level-crossing of the phases.

% -------------------------------------------------------------------
% -------------------------------------------------------------------

We observe the following features.
$(i)$~For low magnetic fields ($0\leq B/J < 0.27$), the ground state is a spin spiral phase.
In this phase, the spins rotate about an axis in the XY plane with a pitch angle
such that the wavevector $Q = 2\pi/\lambda \sim D/J$.
With increasing temperature, thermal fluctuations randomize the spins,
and above a critical temperature the state becomes a paramagnet. 
$(ii)$~At intermediate field strengths ($0.27 \leq B/J \approx 0.65$) 
a skyrmion phase is obtained as the ground state. 
In this phase, the spin components $S_{ix}$ and $S_{iy}$ 
form an array of two dimensional vortices.
This phase shows a periodic arrangement of individual skyrmions, 
and is known as the skyrmion crystal (SkX) phase.
With increasing temperature, the periodic arrangement of skyrmions 
is lost to thermal fluctuations, and above a critical threshold, 
leads to a skyrmion liquid (SkL) phase. 
The skyrmions in this phase are elongated, strongly distorted, and 
distributed randomly in a background of field-polarized spins.
Unlike the low temperature SkX phase, 
the SkL phase survives till large temperature ($T \approx J$).
The stability of the SkL phase is attributed to 
the topological nature of individual skyrmions.
$(iii)$~For high magnetic field values ($B/J \geq 0.65$), 
the skyrmion phase gets suppressed completely, 
and a ferromagnetic phase polarized along the field direction is obtained. 
Further increasing the magnetic field doesn't change the symmetry of this state.
This polarized phase begins to randomize at large temperature ($T \approx J$) regime. 
A quantitative characterization of these different phases is obtained from the
study of multiple observables as detailed below.

{\bf Magnetization:} 
The behavior of magnetization $M$ as a function of 
increasing magnetic field is shown in Fig. \ref{gs}(b). 
In the ground state, $M$ is vanishingly small 
in the absence of an external field. 
It increases slowly at small fields, with an almost linear dependence on $B$. 
At $B\approx 0.27J$, there is a jump in the value of the magnetization, 
signaling a discontinuous magnetic phase transition. 
In the field range $ 0.27J \lesssim B \lesssim 0.65J$,
the magnetization increases with increasing external field. 
As $B \geq 0.65J$, $M$ saturates to its maximum value
indicating the appearance of a fully polarized phase.
With increasing temperature, 
thermal fluctuations randomize the spin orientations
and as a result reduce $M$.

% -------------------------------------------------------------------
\begin{figure*}[t]
\centering
\includegraphics[width=13.5cm,height=7.3cm]{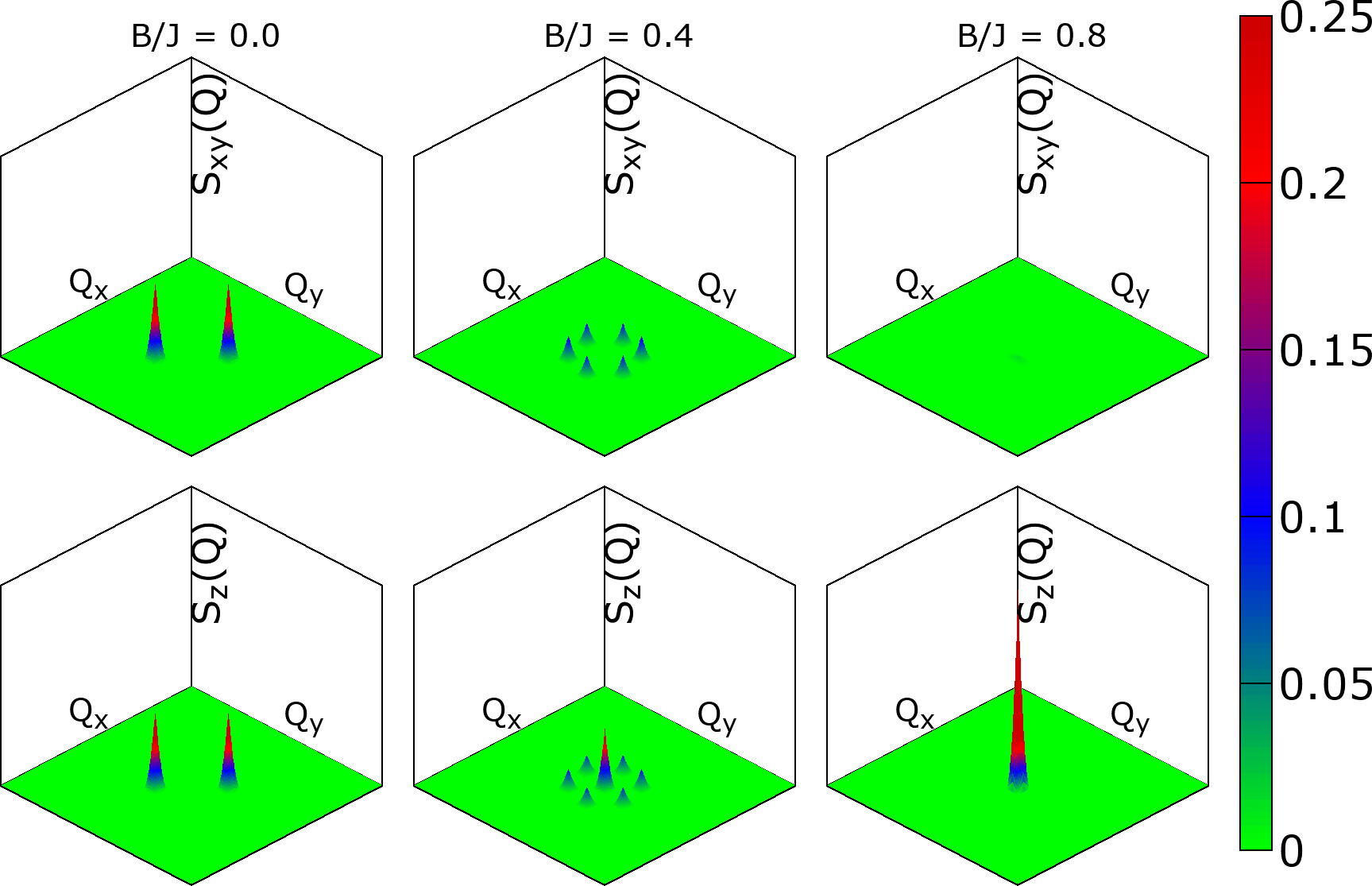}
\caption{\label{sf1}
Color online: Spin structure factor components 
$S_{xy}({\bf Q})$ and $S_{z}({\bf Q})$ in the ground state
calculated for $Q_{x}$, $Q_{y}$ $\in$ $[-\pi,\pi]$ 
at indicated Zeeman field values for $D = J$.
}
\end{figure*}
% -------------------------------------------------------------------

{\bf Structure factor:} A detailed understanding of the multiple
magnetic phases is provided by the spin structure factor, $S(\bf{Q})$,
which quantifies long range magnetic order in terms of prominent peaks 
(or dominant weight) in the momentum space. 
To isolate the effect of the longitudinal field on the spin correlations, 
we have calculated the structure factor components 
$S_{xy}({\bf Q})$ and $S_{z}({\bf Q})$ separately by using the 
$xy$-components and the $z$-components of the spins, respectively. 
We observe the following features (See Fig. \ref{sf1}),
$(i)$~In the spiral (low field) phase, both $S_{xy}({\bf Q})$ and $S_{z}({\bf Q})$ 
show two prominent Bragg peaks (seen as dominant weights in the color map) 
in the $Q_{x}$ and $Q_{y}$ plane. The two ordering momenta are not
independent, but related by symmetry. This non-collinear long range ordered
phase is specified by one wavevector (1-{\bf Q} state).
$(ii)$~In the SkX phase, $S_{xy}({\bf Q})$ shows 6-peaks 
in the $Q_{x}$ and $Q_{y}$ plane -- of these {\bf Q}'s, 
three are independent and for each peak there is another one related by symmetry. 
This state can be understood as a linear superposition of three spiral phases 
(each represented by two symmetry related wave-vectors), 
and is termed a 3-{\bf Q} state. The sharp peaks in these
phase denotes a near-perfect close packed ordering of the skyrmions. 
$S_{z}({\bf Q})$ shows 7-peaks in the $Q_{x}$ and $Q_{y}$ plane 
with the additional peak corresponding to ${\bf Q} = 0$ 
due to the uniform magnetization along the longitudinal direction.
$(iii)$~In the fully polarized ferromagnetic phase, 
$S_{z}({\bf Q})$ shows a prominent peak at ${\bf Q} = 0$,
and $S_{xy}({\bf Q})$ shows no observable weight.
$(iv)$~The SkL phase, which stabilizes 
at finite temperature above the SkX phase, 
shows very weak, and diffused features in $S_{xy}({\bf Q})$
in the $Q_{x}$ and $Q_{y}$ plane (not shown in the figure). 
This is due to the effect of thermal fluctuations on skyrmion configurations 
resulting in distorted skyrmions which lack long-range order 
and only show short range correlations.

\begin{figure}[b]
\centering
\includegraphics[width=6.0cm,height=4.5cm]{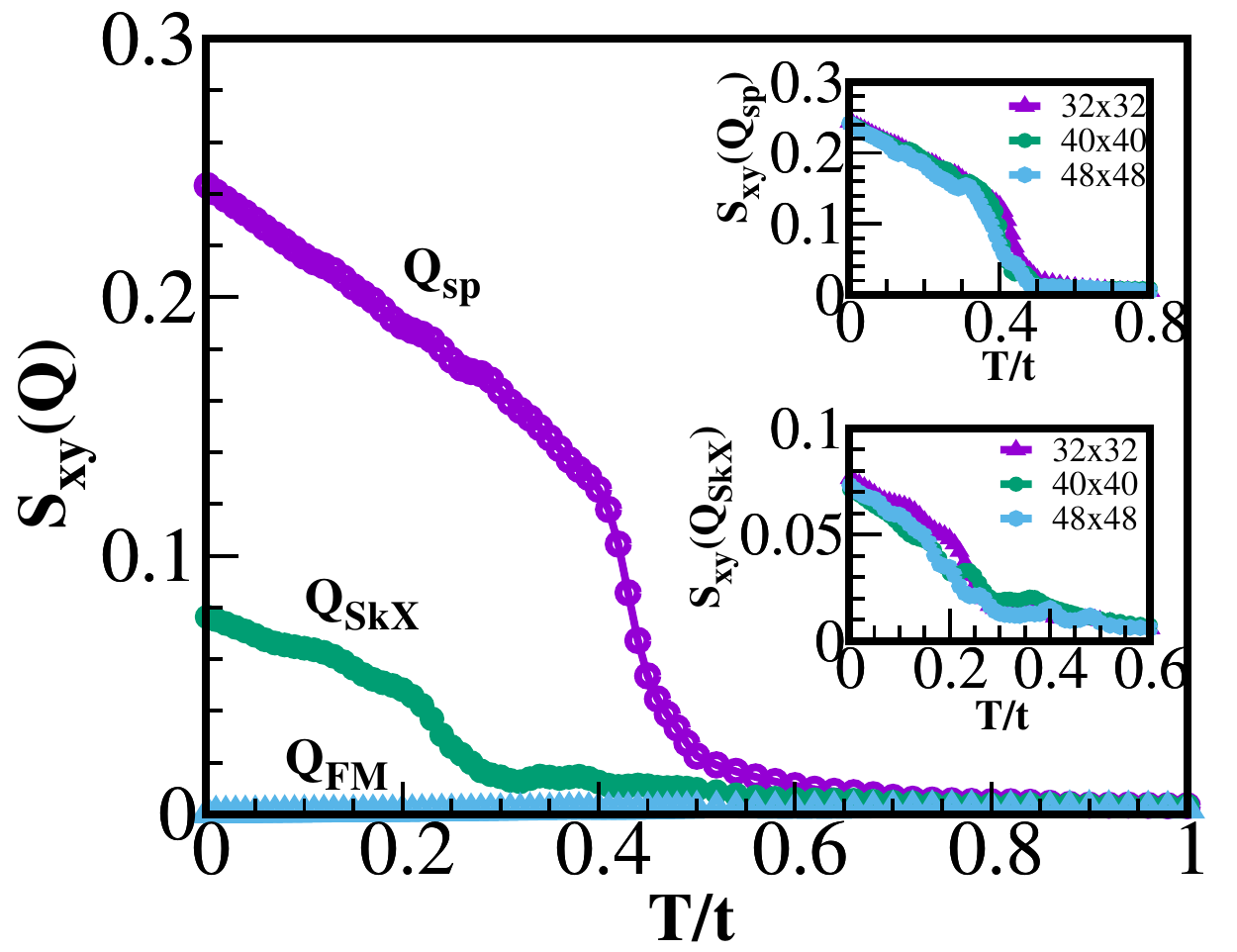}
\caption{\label{sf2}
Color online: Variation of the peak of the structure factor component 
$S_{xy}({\bf Q})$ with temperature for $B/J = $ 0.0, 0.4 and 0.8, 
corresponding to the spiral, skyrmion and the field-polarized phases respectively. 
The point of inflection separates the low-temperature phase 
from the high-temperature phase and represents the critical temperature ($T_c$)
for the respective phases. The insets show the variation of the structure factor
peak for the spiral and skyrmion crystal phases for different system sizes.
}
\end{figure}

Further we look into the variation of the peak of the structure factor component
$S_{xy}({\bf Q})$ with increasing temperature (see Fig.\ref{sf2}).
We observe that the structure factor weight reducing monotonically 
with increasing temperature across all phases.
This is expected as increased thermal fluctuations randomize the magnetic phases,
thus reducing the long-range correlation and eventually resulting in the loss of 
long-range order. The structure factor weight shows a stronger suppression 
beyond the critical temperature. We estimate the critical temperature, $T_c$, as 
the point of inflection of structure factor weight variation with increasing temperature. 
Our results indicate that the spiral phase shows a sharp reduction of the weight
beyond $T_c$ in contrast to the skyrmion phase, where the reduction is milder.

% -------------------------------------------------------------------
% -------------------------------------------------------------------

\begin{figure}[t]
\centering
\includegraphics[width=7.0cm,height=4.8cm]{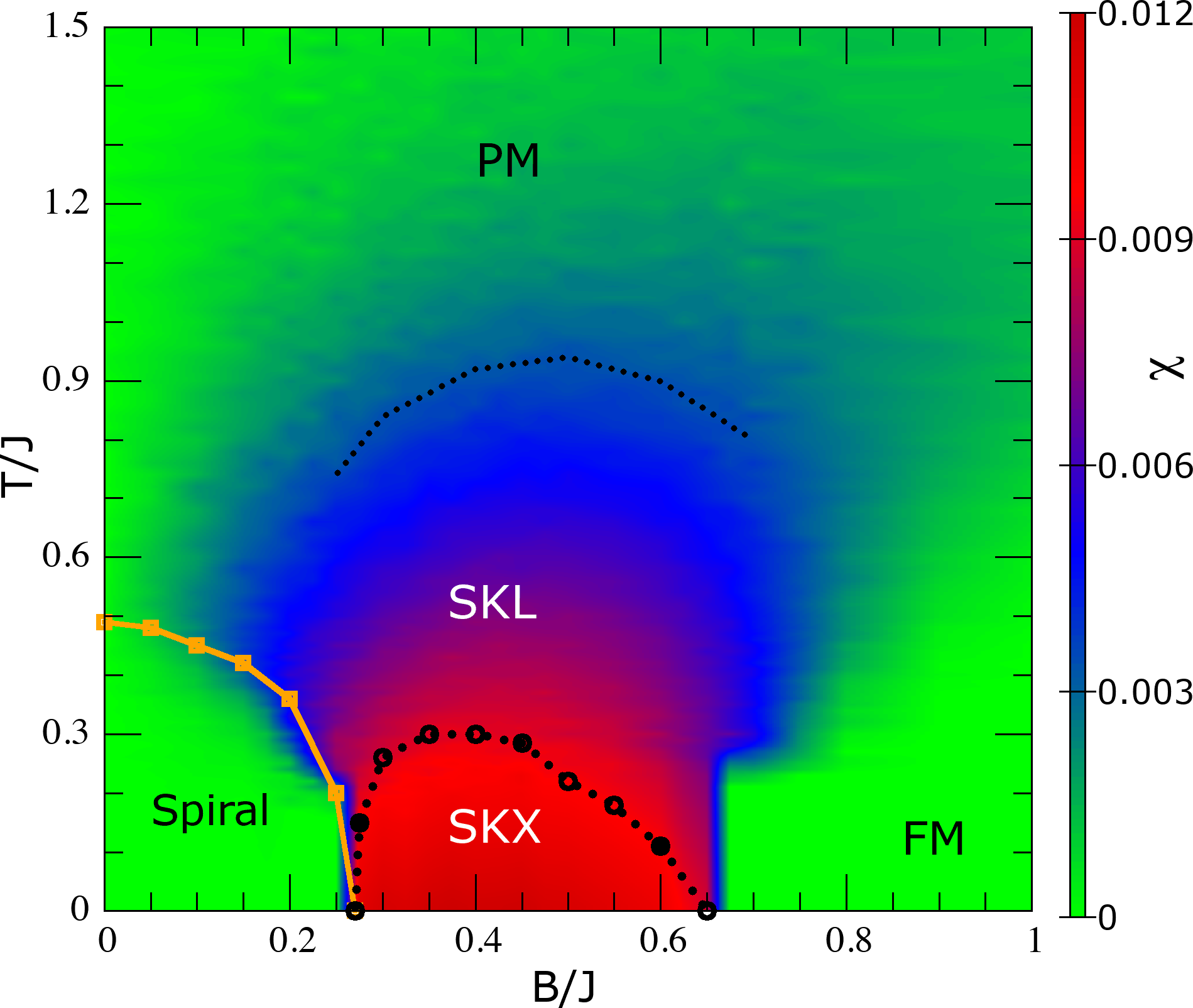}
\caption{\label{spin-chir}
Color online: Spin chirality, $\chi$, obtained with our Monte-Carlo simulation
for different Zeeman field values and temperatures at $D = J$. 
$\chi$ is non-zero in the skyrmion phase. 
Further, we highlight the different phases of our model here. 
At low temperature, thermal fluctuations lead to metastable states, 
resulting in narrow regions over which the phase transitions occur.
The continuous line corresponds to a thermal phase transition 
from the spiral phase to a paramagnet brought by thermal fluctuations, 
and is inferred from the variation of the structure factor weight 
with increasing temperature (see Fig. \ref{sf2}). 
The dotted line separating the SkX phase from the SkL phase refers
to a crossover highlighting the loss of long-range order in the SkL phase,
but finite $\chi$ in both phases.
The dotted line separating the SkL phase from the
paramagnetic phase is also a crossover as indicated by the variation
of Hall conductivity with increased temperature (see Fig. \ref{cond1}(b)).
Further, the SkL phase is separated from the fully polarized FM phase 
by a crossover regime.
}
\end{figure}

{\bf Spin chirality:} One of the most interesting characteristics of 
complex non-coplanar spin textures is the non-zero chirality associated with them. 
Various spin texture are stabilized in the current system due to 
the interplay of the exchange interaction, the DM interaction and 
the applied magnetic field. To quantify the non-coplanarity of these spin textures 
we look into the scalar spin-chirality $\chi$.
Our calculation of $\chi$ gives the following results (see Fig. \ref{gs}(c) and Fig. \ref{spin-chir}).
$(i)$~The low-temperature spiral phase is comprised of spins 
arranged in a non-collinear, but co-planar configuration. 
Accordingly, the $\chi = 0$ in this phase.
With increased temperature thermal fluctuations gives rise to a small,
non-zero value of $\chi$ in this regime.
$(ii)$~With the onset of the SkX phase at $B/J\approx 0.27$, 
the chirality increases discontinuously to a large value, 
reflecting the non-coplanarity of the spin configurations in this phase. 
The non-zero chirality arises entirely from the skyrmionic features. 
Indeed, the spin chirality turns out to be a direct measure of 
the density of skyrmions. The chirality remains almost constant in the SkX phase 
as the density of skyrmions does not change. 
Upon the melting of the SkX into a SkL phase 
with increasing temperature $T \gtrsim T_c$,
the density of skyrmions decrease rapidly. 
Consequently, the chirality exhibits a similar behavior. 
However, it is non-zero even at large temperature regime ($T \approx J$). 
$(iii)$~As $B/J \geq 0.65$ at low temperature, 
chirality reduces due to increased polarization of the spins.
For $B/J \geq 0.8$, $\chi = 0$, showing the fully polarization of the spins 
and the complete loss of non-coplanarity. 
At higher temperature, increased thermal fluctuations in this regime 
may lead to non-zero spin-chirality.

% -------------------------------------------------------------------
% -------------------------------------------------------------------

To summarize, Hamiltonian (\ref{equ:ham-class}) exhibits a sequence of
non-collinear magnetic ground states in an external magnetic field. 
In the absence of an external field, 
the interplay between the Heisenberg and DM interactions
result in a spiral ordering of the localized moments. The spins rotate about 
an axis in the XY plane with a pitch angle that depends on the strength of
the DM interaction. The periodicity of long range modulation of the spins is 
defined by a single momentum in both the longitudinal and transverse components of the 
static structure factor -- a ``1-${\bf Q}$" state. 
Increasing temperature in this phase leads to a paramagnetic phase
above the critical temperature. 
At small Zeeman fields, the spiral ordering of the spins is maintained, 
but the spins acquire a continuously growing longitudinal component along the applied field.
When the field reaches a critical strength, 
there is a discontinuous transition involving the  nucleation of skyrmions 
and their simultaneous condensation into a self-organized close packed 2D crystal - 
the skyrmion crystal. The size of the skyrmions depends on 
the strength of the DM interaction relative to the Heisenberg interaction. 
The long range ordering of the spins is described by 
the superposition of three spiral orders and is characterized by
three independent peaks in the longitudinal and transverse structure factors.
The non-coplanarity of the spin texture is identified by a non-zero spin chirality.
With increasing temperature, 
the skyrmion crystal melts to a skyrmion liquid phase. 
The skyrmion liquid state -- a thermally disordered phase with isolated
and distorted skyrmions, is marked by a short range order 
arising from the spin textures of individual skyrmions. 
This is reflected in the form of diffused momentum dependence 
of the structure factor. 
The chirality also decreases as the density of skyrmions 
(which are the source of finite chirality) goes down. 
It is important to note that in the skyrmion-liquid phase, 
the suppression of the structure factor is much stronger than 
the suppression of the spin-chirality.  In the ground state,
with increasing magnetic field, the skyrmion crystal phase 
undergoes a continuous transition to a fully spin polarized state.
However, at finite temperature, this will reflect 
in the form of a crossover of phases.
The spin-chirality is zero in this phase.
   
In the skyrmion phase, $\chi$ can also be related to 
the discretized skyrmion number. 
We observe that the discretized skyrmion number is almost constant 
in the skyrmion crystal phase.
With increasing temperature, 
this number decreases continuously in the SkL phase.
The skyrmion number is zero both in the spiral phase 
and the fully field polarized phase. 

\begin{figure}[b]
\centerline{
	\includegraphics[width=6.5cm,height=4.5cm]{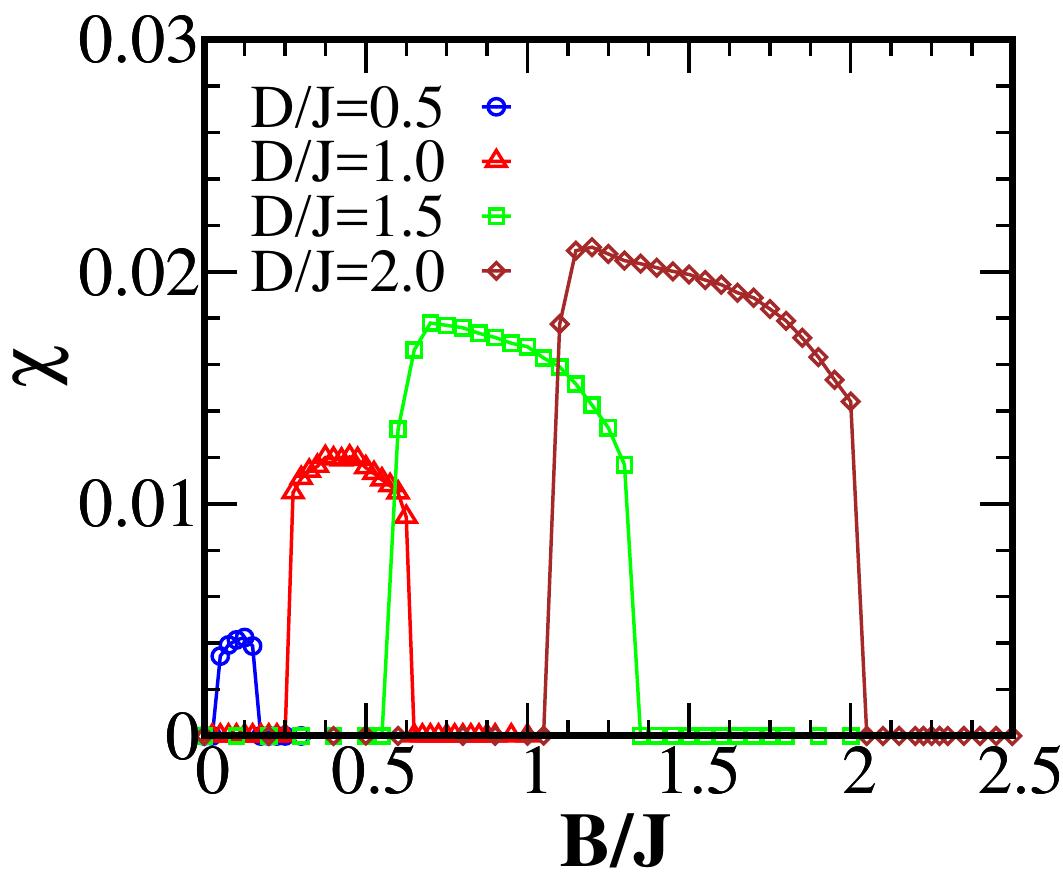}
}
\caption{\label{spin-chir2}
Color online: Ground state spin chirality computed for different Zeeman field values 
and DM interaction strengths.
}
\end{figure}

For larger values of DM interactions, when the skyrmions are smaller and 
the density is larger, the chirality also increases proportionately. 
With varying DM interaction strength, 
we see $\chi$ showing the same qualitative behavior (See Fig. \ref{spin-chir2}).
However, we find the following important differences.
$(i)$~The skyrmion phase is stabilized over a larger Zeeman coupling regime
for increased DM strengths.
$(ii)$~The skyrmion phase has larger chirality values with increasing DM strength,
which suggest that there are more skyrmions (higher skyrmion number) in the SkX phase. 
This also agrees with the previous studies where increasing DM strength leads to 
smaller skyrmion size and hence increased skyrmion number.

%-------------------------------------------------------------------
% -------------------------------------------------------------------

\subsection{Electronic properties}

Coupling to the local moments modifies the transport properties of 
itinerant electrons dramatically. For simplicity, we consider  
a single band of $s$-electrons interacting with the magnetic ordering 
via a Kondo coupling between the electron spin and the local moments.
The dynamics of the electrons is fast compared to that of the localized
classical spins. Consequently, at short time scales, the electrons effectively 
move in a static, but spatially varying magnetic field. 
Each local moment, ${\bf S}_i$ acts as a local magnetic field 
whose action on the spin magnetic moment of the itinerant electrons 
${\bf s}_i$ is described by a Kondo-like interaction 
$J_K{\bf s}_i\cdot{\bf S}_i$. 
In comparison, the Zeeman energy due the external magnetic field coupled to the 
spin of the electron is small and shall be neglected. In the following, we 
discuss the effects of the different field induced spin textures on the 
energy dispersion and conductivity of the electrons.

% -------------------------------------------------------------------
\begin{figure}[t]
\centerline{
	\includegraphics[width=8.4cm,height=4.5cm]{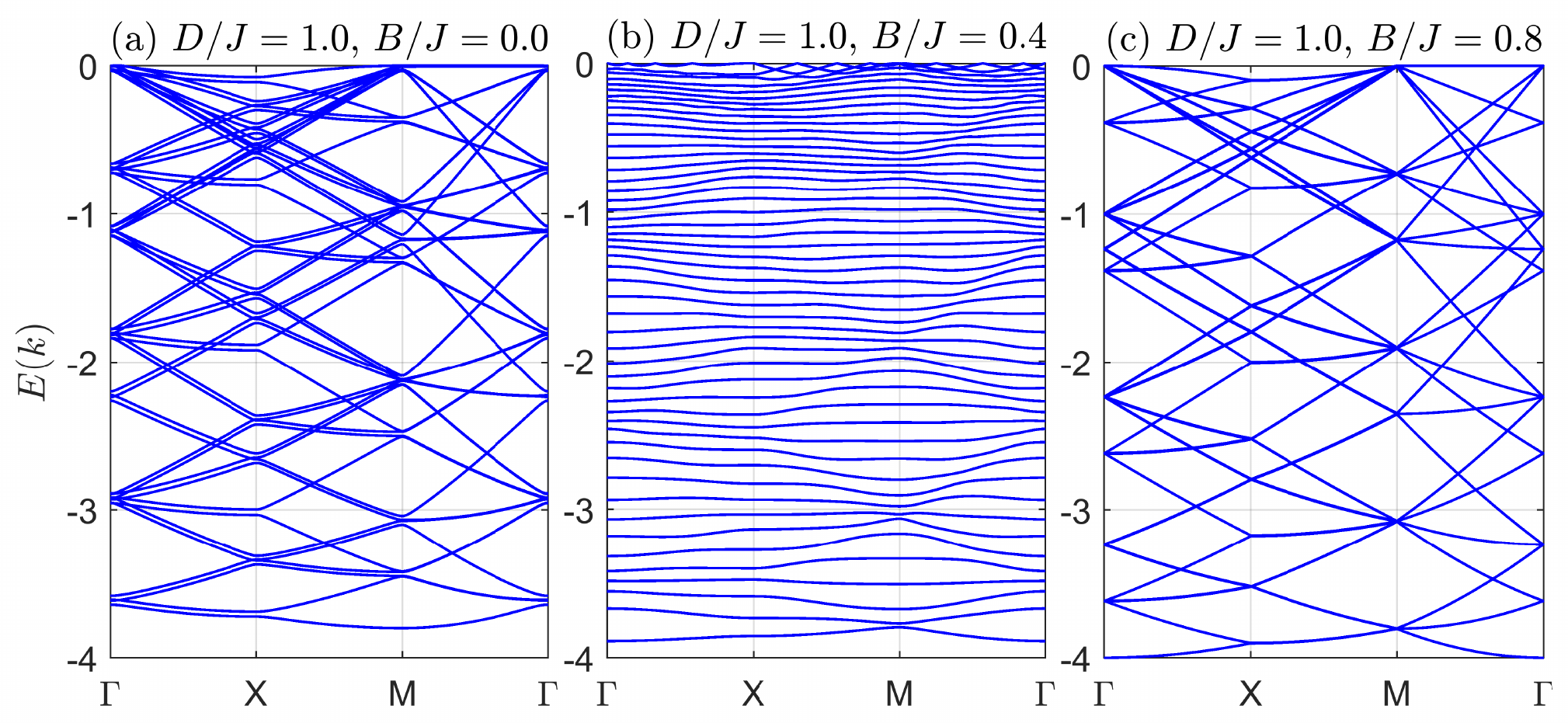}
}
\caption{\label{band-str}
Color online: Band structure of conduction electrons moving on the spin backgrounds of
spiral, skyrmion and ferromagnetic phases for indicated values of DM interaction 
and magnetic field.
}
\end{figure}
% -------------------------------------------------------------------

{\bf Band structure :}
In the absence of an external field, the electron band structure consists of
a single band with 2-fold spin degeneracy, and a band width $w=8t$. 
There exist van Hove singularities at ${\bf k} = (0,\pm\pi)$ and $(\pm\pi,0)$ 
where any interaction effects are maximized. A coupling to the spin texture 
increases the size of the unit cell in accordance with the magnetic unit cell. 
The Brillouin zone (BZ) is proportionately reduced and the bands are folded
into the first BZ. A non-zero $J_K$ lifts the spin degeneracy and the energy bands 
for electrons with spins parallel and anti-parallel to the local moments 
are shifted downwards and upwards, respectively. For sufficiently strong
$J_K (> 8t)$, the spin parallel and anti-parallel bands 
are completely separated by a gap. In the limit of $J_K \gg t$, 
the electron spins are completely aligned with the local moments 
and Hamiltonian (\ref{equ:ham-elec}) reduces to an effective tight-binding model,
\begin{equation}
 {\mathcal{\hat{H}}_e}= -\sum_{\avg{i,j},\sigma}t_{ij}^{eff}(d_{i}^\dagger d_{j}+\mbox{H.c.}).
\end{equation}
Where

\begin{equation}
t_{ij}^{eff}=t e^{ia_{ij}}\cos\frac{\theta_{ij}}{2},
\label{eq:teff} 
\end{equation}

is the effective hopping matrix for the spin-parallel electrons between sites $i$ and $j$ and
\begin{equation}
a_{ij}=\arctan\frac{-\sin(\phi_i-\phi_j)}{\cos(\phi_i-\phi_j)+\cot\frac{\theta_i}{2}\cot\frac{\theta_j}{2}}
\end{equation}
is the phase factor and $\theta_{ij}$ is the angle difference 
between two localized spins $\mathbf{S}_i$ and $\mathbf{S}_j$.
The spin anti-parallel electrons are described by 
a similar effective tight binding model with a different $t_{ij}^{eff}$ 
and the two sectors are completely decoupled. 
The detailed effects of the magnetic orders 
in the different field induced phases are discussed below. 
For illustrative purpose, we have chosen an unphysically high value of 
the DM interaction to stabilize the magnetic order with short periodicity. 
This enhances the effects of electron coupling to local moments, 
allowing us to demonstrate them more effectively. Similar effects persist at smaller
(and physically relevant) DM interaction strengths, but at a much smaller scale.

In this limit, the effective magnetic field produced by the spin texture 
couples directly to the charge of the electrons analogous to quantum Hall systems. 
The electron energy bands are modified 
depending on the nature of the underlying magnetic order. 

To study the topological nature of the bands, 
we calculate the Chern number associated with each band as 
$C_{n} = \frac{1}{2\pi} \int_{\mathrm{BZ}}\Omega_{n}^{(z)}({\bf k}) \mathrm{d}^{2}{\bf k}$
where, the Berry curvature is given by
$\Omega_{n}^{(z)}({\bf k}) = 
\partial_{k_{x}} A_{n}^{(y)}({\bf k}) - \partial_{k_{y}}A_{n}^{(x)}({\bf k})$ 
with 
${\bf A}_{n}({\bf k}) = 
-\mathrm{i} \langle u_{n}({\bf k})|\nabla_{{\bf k}}|u_{n}({\bf k}) \rangle$
as the Berry connection calculated from the eigenvectors $u_{n}({\bf k})$ 
with eigenvalues $E_{n}({\bf k})$ of the Hamiltonian~\ref{equ:ham-elec}.
We focus on the band structure corresponding to different magnetic phases.

% -------------------------------------------------------------------
\begin{figure}[b]
\centerline{
	\includegraphics[width=8.4cm,height=4.6cm]{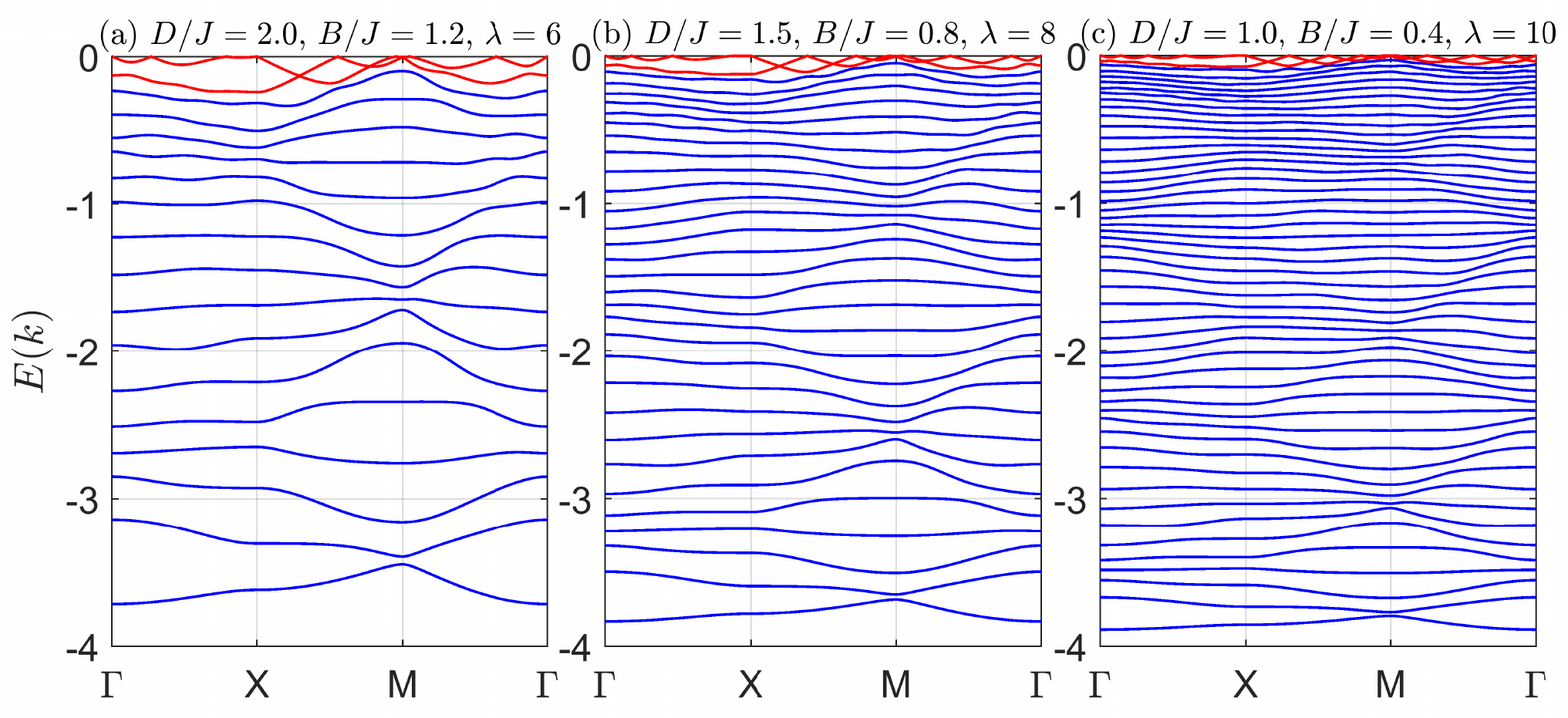}
}
\caption{\label{band-str2}
Color online: Band structure of the conduction electrons on the skyrmion crystal background
for indicated DM strengths and Zeeman coupling values. The bands, drawn in blue are
separated from each other by a gap and have Chern number $C=-1$, 
whereas the bands drawn in red (two highest energy bands) are touching each other 
with Chern number equal to the sum of Chern numbers of the lower bands with opposite sign.
}
\end{figure}
% -------------------------------------------------------------------

% -------------------------------------------------------------------
\begin{figure}[t]
\centerline{
	\includegraphics[width=4.1cm,height=4.5cm]{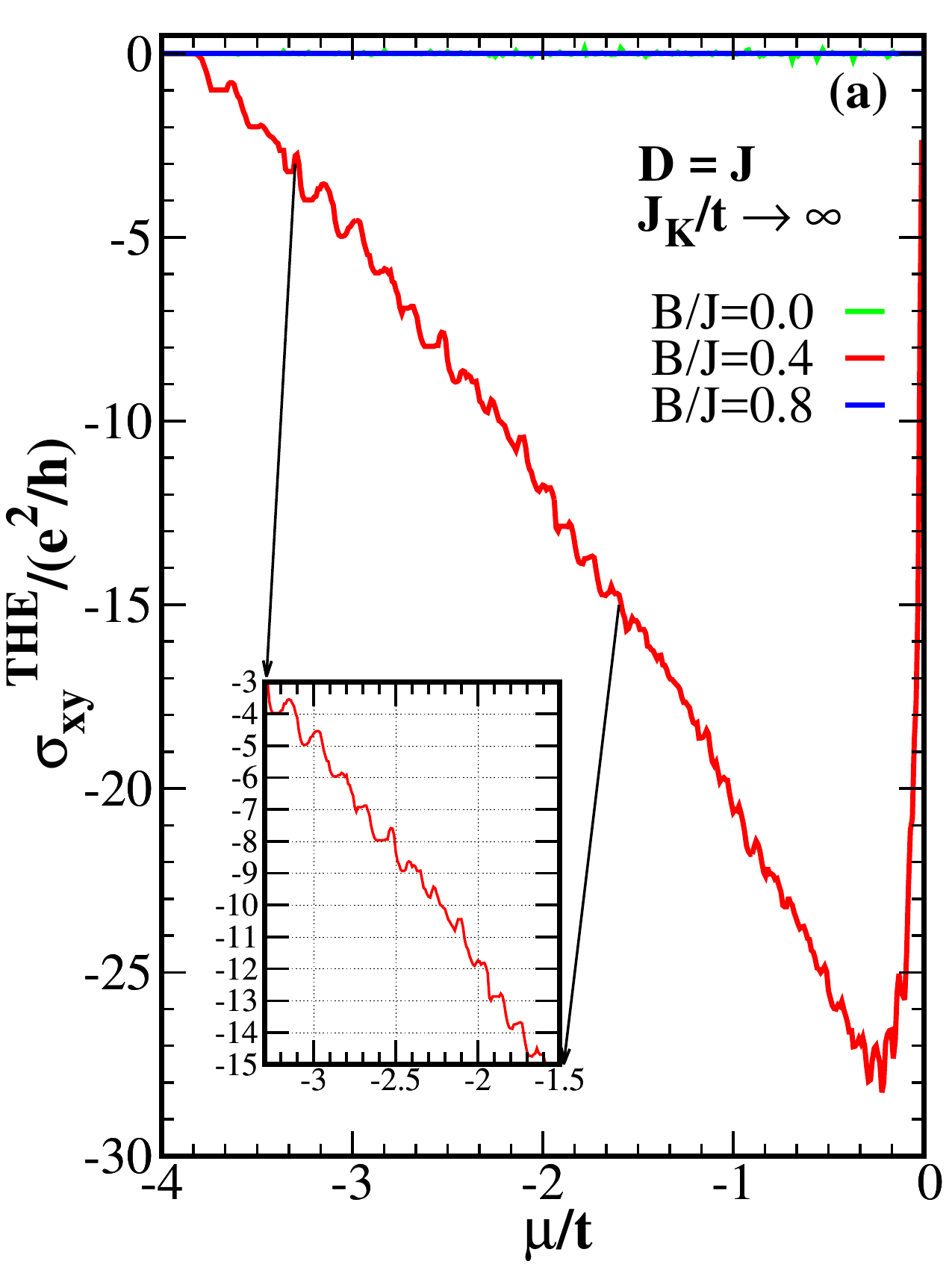}
	\includegraphics[width=4.1cm,height=4.5cm]{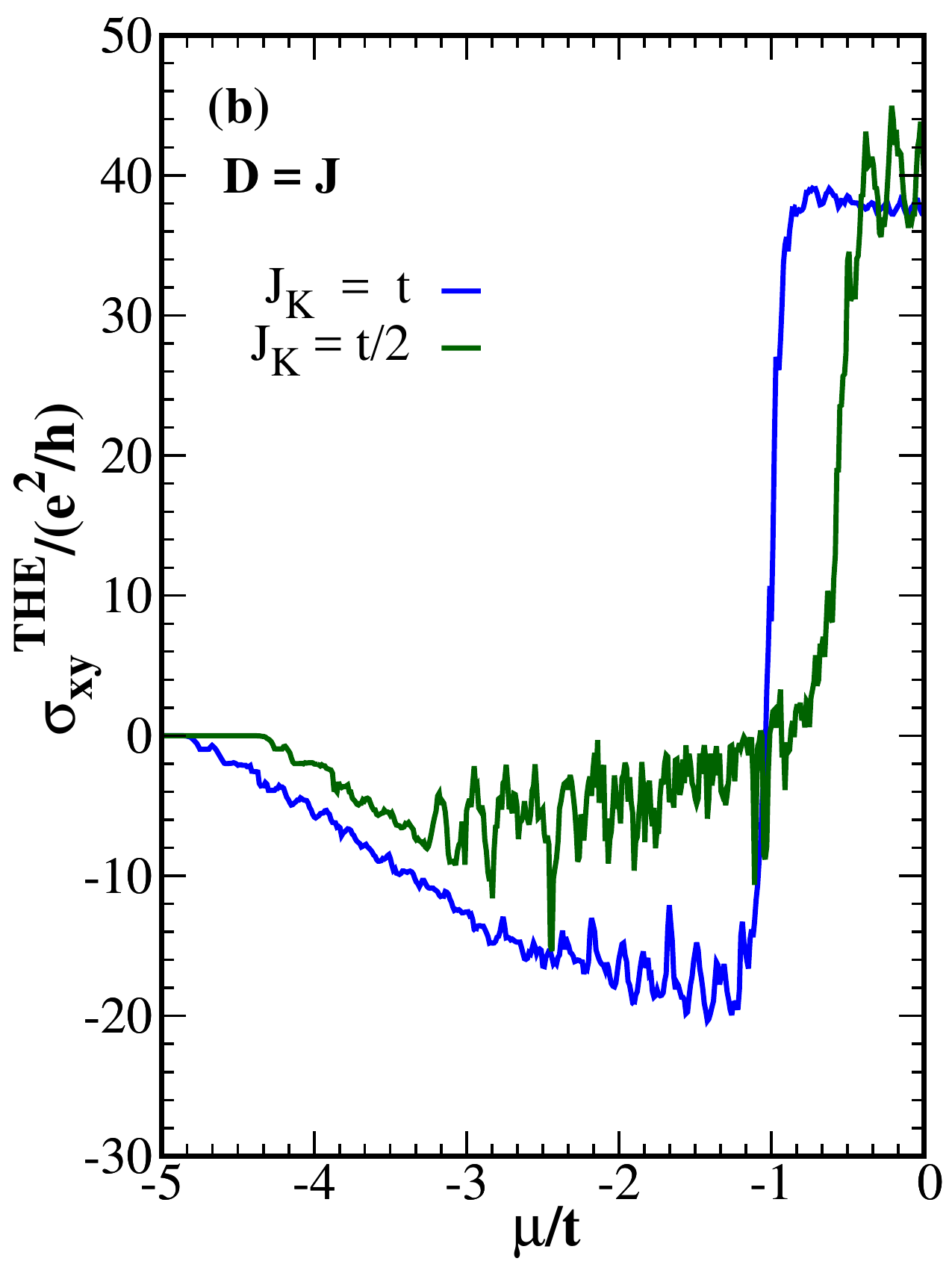}
}
\centerline{
	\includegraphics[width=4.0cm,height=4.5cm]{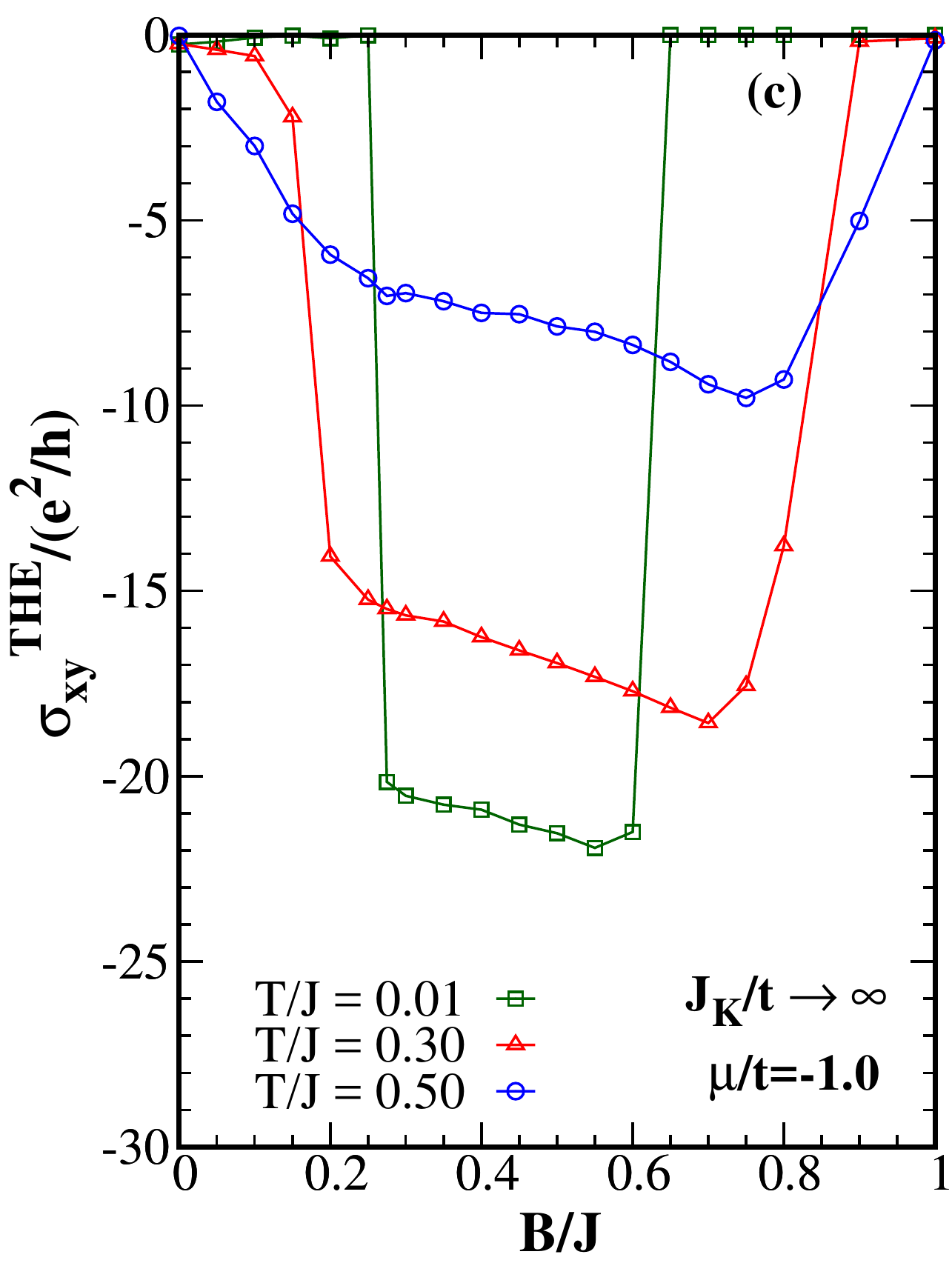}
	\includegraphics[width=4.0cm,height=4.5cm]{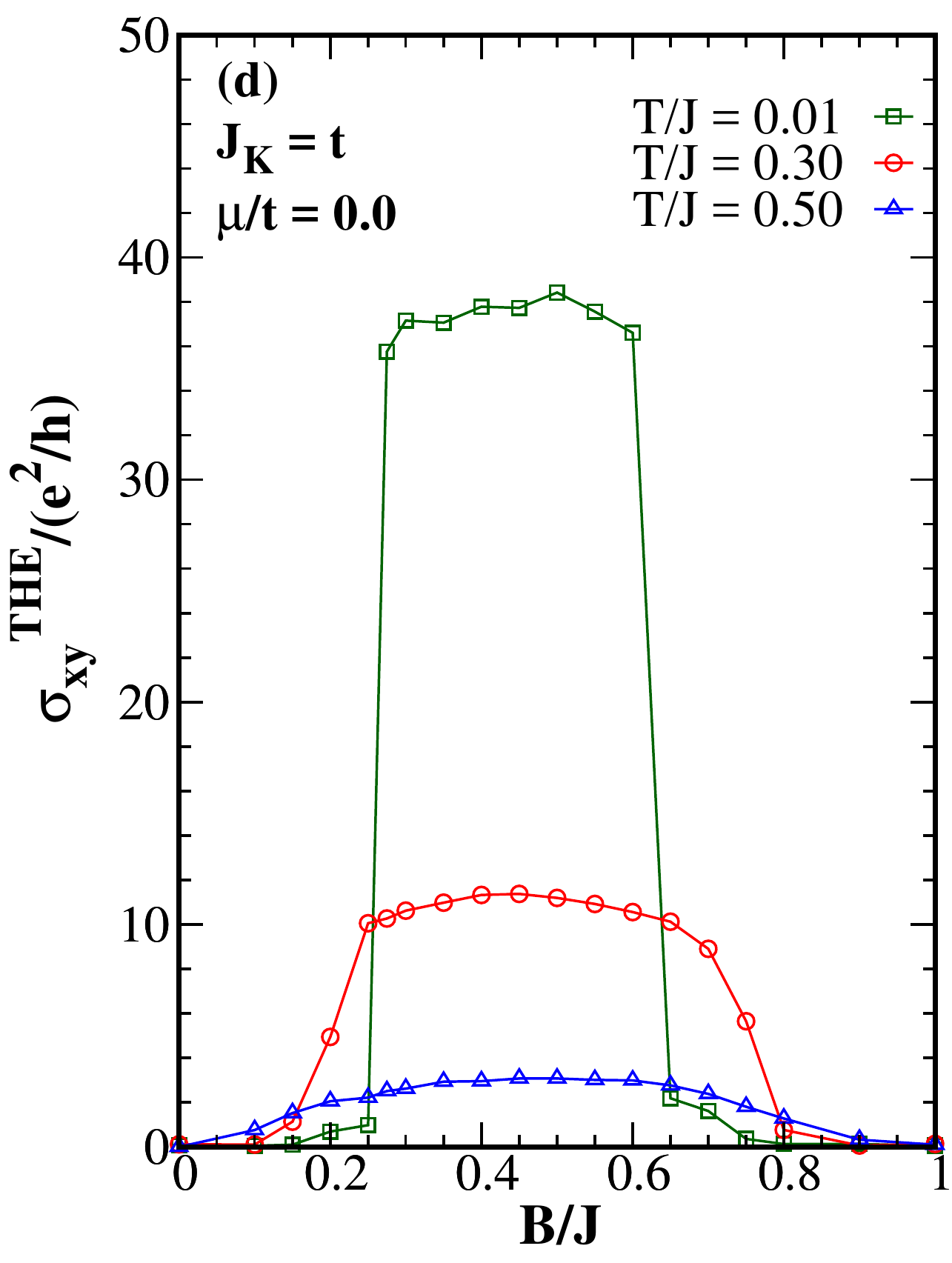}
}
\caption{\label{cond}
Color online: (a)-(b)~Behavior of the topological Hall conductivity
with varying chemical potential in the ground state for $D = J$. 
Panel (a) shows the dependence of  $\sigma_{xy}$  on the chemical potential in 
the strong coupling limit ( $J_K \rightarrow \infty$), for representative values of 
the applied field corresponding to spiral, SkX and ferromagnetic ground states. 
Panel (b) shows the effects of the strength of Kondo coupling on the variation
of topological Hall conductivity with chemical potential for the SkX ground
state. Results for three representative  values of $J_K/t$ are shown,
corresponding to strong, intermediate and weak coupling strengths. For
intermediate to weak coupling, the topological Hall conductivity is finite at $\mu=0$,  
which is in a sharp contrast to the behavior for strong coupling. 
(c)-(d)~Variation of the topological Hall conductivity at fixed chemical potential 
with changing magnetic field at different temperatures. 
Panel (c) corresponds to parameters $\mu = -t$ and $J_K \rightarrow \infty$, whereas, 
panel (d) corresponds to $\mu = 0$ and $J_K = t$.
In the low temperature regime, $\sigma_{xy}$ is finite only in the skyrmion phase
and shows sharp jumps across the phase boundaries. 
For intermediate and high temperature regimes, $\sigma_{xy}$, 
increases rather continuously from the spiral to the skyrmion phase,
and vanishes only deep in the polarized phase.
While the conductivity at finite $J_k$ shows qualitatively similar behavior 
to the $J_K \rightarrow \infty$ limit at low temperatures, the thermally induced 
$\sigma_{xy}$ for spiral and ferromagnetic ground states are more pronounced 
for strong coupling strength.
}
\end{figure}
% -------------------------------------------------------------------

\noindent (i)\underline{\it{Spiral phase}}:  The periodicity of 
the magnetic ordering at $D=1.5J$ is observed to be $\lambda =8$. 
This translates to a magnetic unit cell of size $\lambda\times\lambda$. The 
Brillouin zone (BZ) is reduced to $-{\frac{\pi}{\lambda}} \leq k_x,k_y < {{\frac{\pi}{\lambda}}}$. 
The band structure consists of 2$\lambda^2$ bands 
that are completely separated for $J_K > 8t$. 
Fig.\ref{band-str} shows the band dispersion for the spin-parallel bands 
in the strong coupling limit along a high symmetry path in the 1st BZ, 
calculated using the effective tight binding model. 
For clarity of presentation, only the bands with $E({\bf k}) \leq 0$ are shown. 
The bands are modified from their non-interacting limit, but 
retain the large dispersion. No opening of bands gaps is observed.

\noindent (ii)\underline{\it{Skyrmion Crystal phase}}: The effect of 
the spin texture is most pronounced in the skyrmion crystal phase, 
and has been studied in recent past with {\it ideal} skyrmions being
constructed on different lattices~\cite{Nagaosa_SkX_THE_2015,Mertig_SkX_THE_2017}. 
In the current work, we present results for interaction driven, 
{\it self consistently generated} skyrmion crystal phase. 
For the non-coplanar spin ordering of the skyrmions, 
the local moments around a plaquette subtend a finite solid angle
at the center due to the  spatially varying spin texture. 
This results in a finite Berry phase when an electron hops around a plaquette. 
In the strong coupling limit, the phase of the effective hopping, 
$a_{ij}$, can be associated with a vector potential 
acting on the itinerant electrons, analogous to quantum Hall systems. 
The flux of the corresponding fictitious magnetic field 
through each plaquette is given by $\Phi = (n_{sk}/\lambda^2)h/e$. 
The dispersion of itinerant electrons is strongly affected by this Berry phase. 
The bands get narrower. For the lower bands, 
a gap opens up between successive pairs of bands and 
each band acquires a finite Chern number (see fig. \ref{band-str2}). 
However, the two uppermost bands are not complete gapped out;
instead they touch each other at several ${\bf k}-$points in the BZ. 
Drawing on the analogy with quantum Hall systems, 
the bands can be described as dispersive Landau levels. 
For the (lower) block shown, each band carries a Chern number $C= -1$. 
Close to the van Hove singularity (VHS), two bands touch each other,
and they carry a large Chern number $C = \lambda^2/2 -2$ 
(where $\lambda\times\lambda$ is the size of the magnetic unit cell). 
Above it, the Chern number of each band changes sign. 
This has a dramatic effect on the behavior of the transverse conductivity, 
as discussed below.

The energy bands in the skyrmion crystal phase 
for different values of DM interaction are plotted in Fig.\ref{band-str2}, 
along the same path in the BZ. The plotted bands correspond to 
half of the number of bands (symmetric about $\mu =0$), 
where conduction-electron spins are aligned with the local moments.
Smaller values of $D/J$ result in larger skyrmions.
This leads to larger magnetic unit cells; 
correspondingly, the number of bands in the BZ increases. 
However, the qualitative features remain unchanged.
The lower bands are separated from each other via a finite gap, 
have smaller bandwidths, and each carry a Chern number of $C= -1$; 
whereas the top two bands near $\mu = 0$ touch each other, 
and have a combined Chern number, $C= \lambda^2/2 -2$.

\noindent (iii) \underline{\it{Fully polarized phase}}: In the fully polarized phase, 
the electrons simply acquire a magnetic potential energy 
due to the uniform field produced by the local moments. 
The original symmetry of the lattice is restored and the number of
bands reduces to two -- one for each spin. 
The band structure in the reduced BZ is shown in Fig.\ref{band-str}.

% -------------------------------------------------------------------
\begin{figure}[t]
\centerline{
	\includegraphics[width=4.5cm,height=4.0cm]{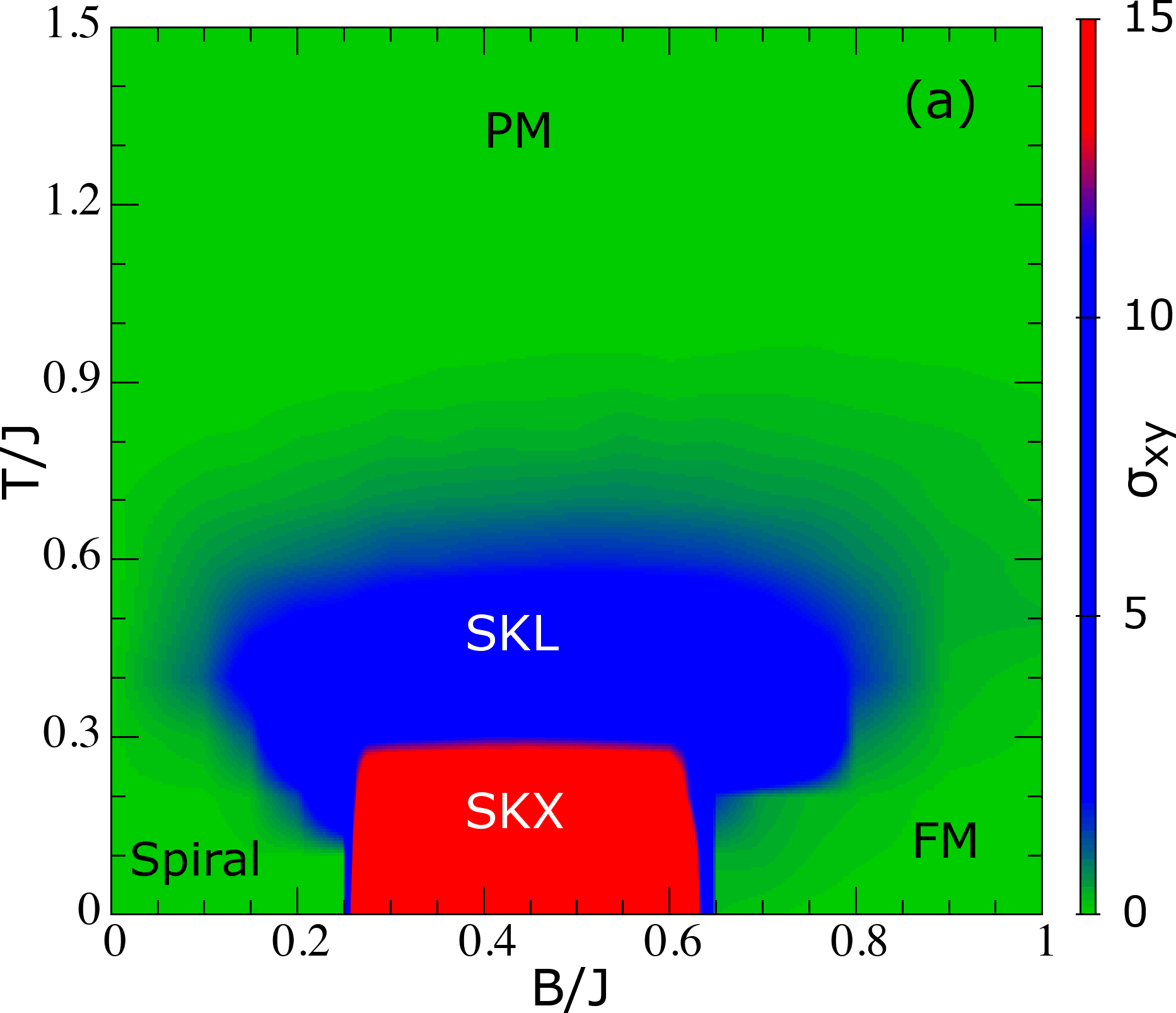}
	\includegraphics[width=4.0cm,height=4.0cm]{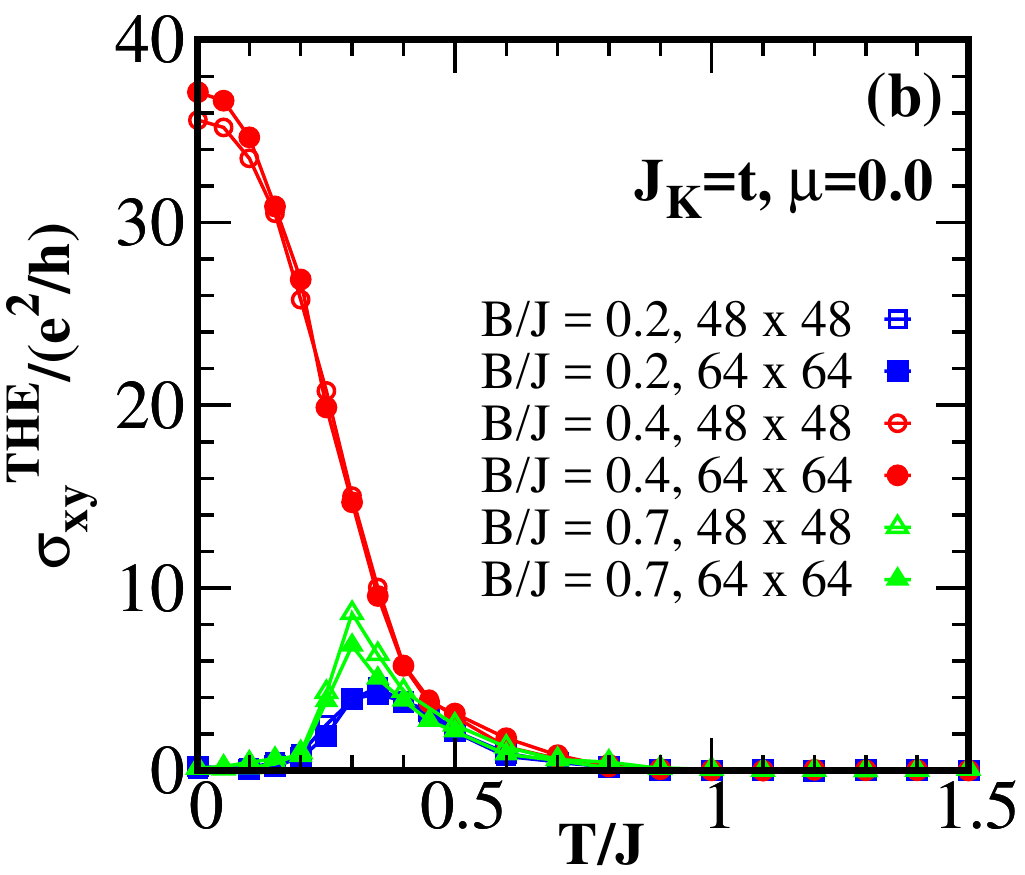}
}
\caption{\label{cond1}
Color online: (a)~Behavior of the topological Hall conductivity with
varying temperature and Zeeman field values for $D = J$.
(b)~Variation of the topological Hall conductivity with temperature 
for $B/J= $0.2, 0.4 and 0.7. 
For $B/J=0.4$, the ground state is deep inside the SkX phase. 
With increasing temperature $\sigma_{xy}$ reduces monotonically
across the SkX and SkL phases, eventually vanishing at $T \approx J$.
The temperature at which $\sigma_{xy}$ reduces by $95\%$ 
of the ground state value is identified as the crossover temperature.
For $B/J =0.2$ and $0.7$, the ground states are in the spiral and
ferromagnetic phases, close to the ground state phase boundary with the SkX phase. 
For both these values of $B/J$, the THE displays a reentrant behavior. 
It increases from zero at low temperatures to a finite value 
for an intermediate range of temperature and
eventually decreases to vanishing values at high temperatures. 
The data for different system sizes show negligible difference, confirming
that the results are not artefacts of finite size effects.
}
\end{figure}
% -------------------------------------------------------------------

{\bf Hall conductivity:} The coupling to local moments modifies the
transport properties of itinerant electrons significantly in metallic magnets. 
The effect is most dramatic in the transverse conductivity, 
especially when the underlying spin arrangement is non-coplanar. 
In a magnetic metal, the Hall resistivity consists of three contributions
\begin{equation}
    \rho_{xy} = \rho_{xy}^\text{NHE} + \rho_{xy}^\text{AHE} + \rho_{xy}^\text{THE}
\end{equation}
where NHE, AHE and THE refer to Normal, Anomalous and Topological
Hall effects, respectively. The AHE appears in metals with a net
magnetization due to spin-orbit coupling. On the other hand, the
THE arises due to the real space Berry phase 
acquired by an electron moving in a non-coplanar spin texture. 
The phenomenon is best understood within
the framework of the effective Hamiltonian (\ref{equ:ham-elec}) in the
strong coupling limit ($J_K\gg t$). In this limit, the Berry phase 
acquired by an electron moving around a closed plaquette results in an
effective flux threading each such plaquette that acts as a fictitious
magnetic field and drives a Hall effect, whose origin is purely geometrical.
In the present work, we focus on isolating the THE contribution to the
transverse conductivity in the various field driven phases. 
Only the SkX and SkL phases exhibit finite THE. 
The spiral phase comprises of spins in a coplanar configuration 
whereas in the fully polarized phase, the spins are all aligned 
parallel to the applied field. The spin chirality vanishes for
both of these configurations and consequently neither exhibit any THE.
Fig. \ref{cond}(a) shows the variation of the zero temperature Hall conductivity 
with changing the chemical potential for the spin aligned electrons 
in the strong coupling limit for the lower half of the bands between the
band minimum and the center of the band (the energies are shifted by
a constant value so the the band extends over the range $[-4t,4t]$). The
conductivity for the upper half is symmetric about the band center. For
the spiral and the fully polarized phases, the transverse conductivity
remains zero, as discussed above. In the SkX phase, transverse conductivity
exhibits a sequence of quantized plateaus analogous to quantum Hall plateaus.
With increasing chemical potential, 
the value of $\sigma^\text{THE}_{xy}$ decreases from zero 
at the band minimum in steps of $e^2/\hbar$, in accordance with the
Chern number $C = -1$ of the lower bands. In contrast to quantum Hall systems,
the conductivity changes continuously, yet non-monotonically between the plateaus. 
This is a consequence of the finite dispersion of the electron energy bands 
(the bands are completely flat for quantum Hall systems). 
Close to the band center (which coincides with the VHS), 
the conductivity increases sharply, reflecting the large and 
positive value of the Chern number of bands close to the VHS. 
The quantization of the Hall plateaus is most pronounced for small 
skyrmion sizes where the lower bands are well separated. For larger
skyrmions, the density of the bands increases and the energy extent of the 
conductivity plateaus decreases proportionately as does the energy separation
between successive plateaus. This makes it difficult to resolve
them in numerical simulations as effects of finite system size and 
fluctuations dominate.

At zero temperature only states below the Fermi energy, 
$E_f$ contribute to the transport.
When the Fermi energy lies within a plateau region, 
there is zero overlap between the current carrying states 
in the sample leading to the absence of backscattering processes. 
Thus the quantized value of $\sigma_{xy}$ signifies 
the absence of backscattering amongst the states. 
If $E_f$ is located within the band gap above any band, 
$\sigma_{xy}$ is proportional to the winding number 
which is the sum of the Chern numbers of the filled bands.
Below the band gap, the Hall conductance decreases, while 
it increases above the band gap. 
This is due to the fact that the sign of the Berry curvature 
is opposite between the two adjacent bands.
This gives rise to a sawtooth shape of $\sigma_{xy}$, 
which becomes more pronounced when the skyrmion size is larger. 
With increasing skyrmion size (reducing the DM interaction), 
the qualitative behavior remains unchanged, 
but the detailed features get harder to resolve.

While the effects of large $J_K (\gg t)$ on ground state electronic transport 
has been extensively studied in the past (see Fig.\ref{cond}(a)), 
the effects of finite temperatures and intermediate values of 
coupling strength between itinerant 
electrons and localized moments ($J_K \approx t$) have remained
largely unexplored. These are specially important for understanding
experimental results, where the interest often lies in room temperature
skyrmions, driven by the potential for practical applications. In the 
following we investigate these effects systematically and demonstrate that
our results for transverse conductivity at finite temperatures agree well
with experimental observations.
Fig.\ref{cond}(b) shows the behavior of the topological Hall conductivity 
with varying chemical potential in the SkX phase in the ground state for finite values of $J_K$, 
together with that for large $J_K$.
The most striking feature of the results for $J_K\lesssim t$ is the 
finite Hall conductivity at $\mu = 0$, which is in a sharp contrast 
to the behavior in the the $J_K \rightarrow \infty$ limit. This is 
because the electrons with spin aligned and anti-aligned to the local
moments are not separated by a large gap for $J_K\lesssim t$. This 
results in multiple bands at the Fermi surface and a finite density of 
states at $\mu = 0$.

% -------------------------------------------------------------------
\begin{figure}[t]
\centerline{
	\includegraphics[width=8.5cm,height=8.0cm]{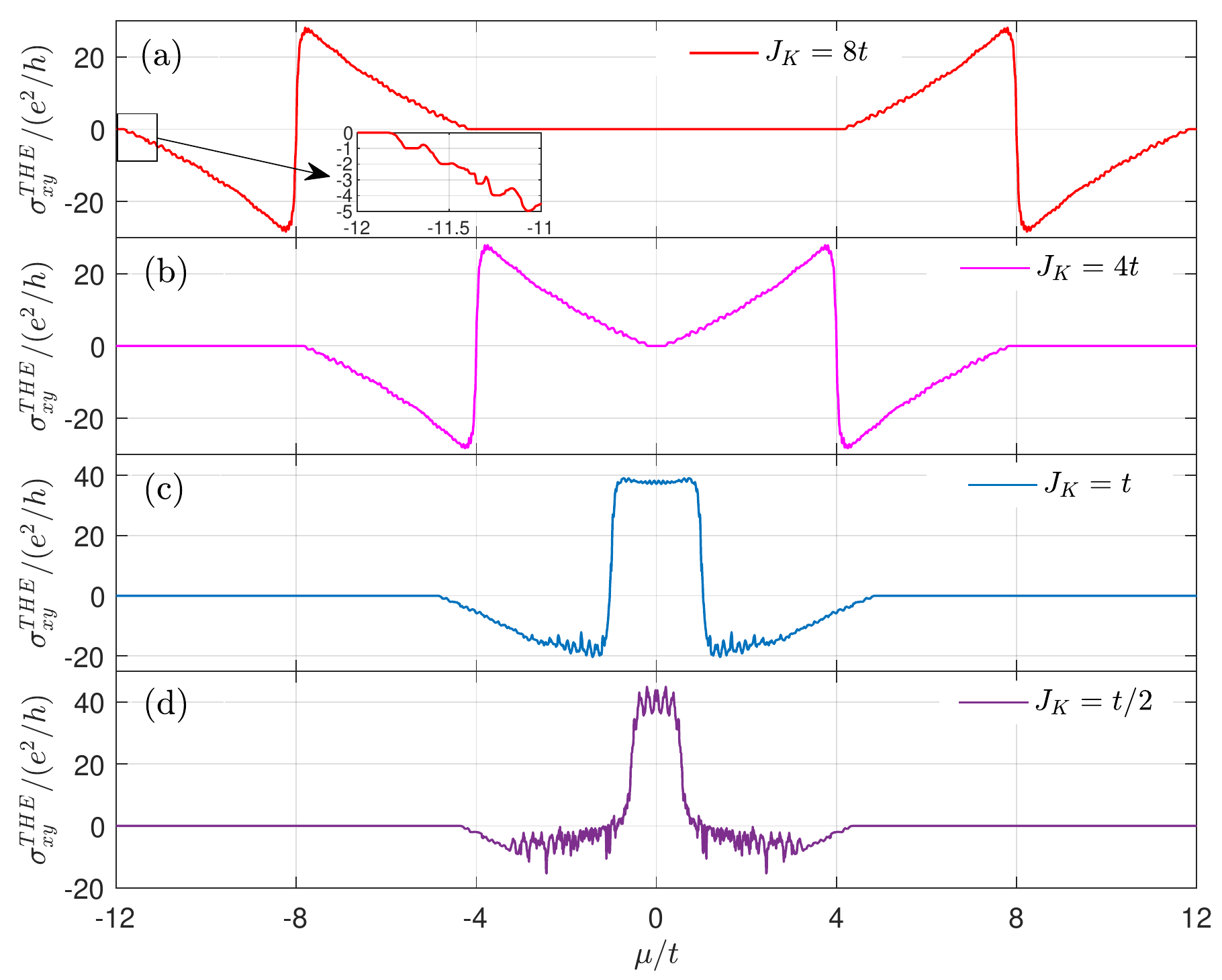}
}
\caption{\label{cond2}
Color online: The topological Hall conductivity as a function of varying chemical potential
in the ground state of the skyrmion crystal phase ($B = 0.4J$) 
for different (finite) values of the Kondo-coupling. 
A finite THE is observed at $\mu=0$ for intermediate to small 
coupling strengths ($J_K \lesssim t$), that is markedly absent at 
stronger couplings.
}
\end{figure}
% -------------------------------------------------------------------

Fig.\ref{cond}(c)-(d) show the variation of the topological Hall conductivity 
with changing magnetic field for different temperatures.
The field dependence of the topological Hall conductivity 
is qualitatively similar for all values of chemical potential 
as long as it is below the van Hove singularity. 
In the ground state ($T/J = 0.01$), we observe that $\sigma_{xy} = 0$ for a weak magnetic field (spiral phase)
and a strong magnetic field (ferromagnetic phase), reflecting the absence of 
Berry phase in the coplanar (spiral) and 
collinear (ferromagnetic) magnetic orderings. 
In contrast, the non-coplanarity of the spin texture in the
SkX phase induces a Berry phase in the conduction electrons which 
is reflected as a finite THE in this phase. 
The Hall conductivity exhibits a discontinuous transition 
at the spiral-SkX and SkX-fully polarized phase boundaries. 
At intermediate temperature, the  Hall conductivity rises monotonically
with increasing field values from well inside the spiral to the skyrmion phase.
Similarly, at high magnetic fields, the transverse Hall conductivity decreases
in magnitude continuously and vanishes deep inside the fully polarized phase. 
Thermal fluctuations induce isolated spin clusters with non-zero chirality in
the spiral and fully polarized ground state phases which results in a finite
$\sigma^{\text{THE}}_{xy}$ inside these  phases.
This behavior is qualitatively similar to the experimental observations~\cite{soumyanarayanan_skm_multi-layer_2017,raju_skm_multi-layer_2019}.
The precise nature of the skyrmion phase is different in the experimental system -- 
in lieu of complete skyrmions, finite-length spin spirals with differently
oriented axes are observed. However, a finite value of the net 
chirality confirms the topological character of the spin texture. 
The observed THE, which is finite in the skyrmion phase, 
exhibits a variation qualitatively very similar to our simulation results. 
This suggests that while a realistic microscopic model of
the experimental system (that faithfully captures its multiple complexities)
is likely to be extremely complicated, 
the simple Hamiltonian~(\ref{equ:ham-class})
represents a minimal model that correctly identifies the principal interactions 
and accompanying thermal effects that give rise to the experimental observations.

% -------------------------------------------------------------------
\begin{figure*}[t]
\centerline{
	\includegraphics[width=12.5cm,height=7.9cm]{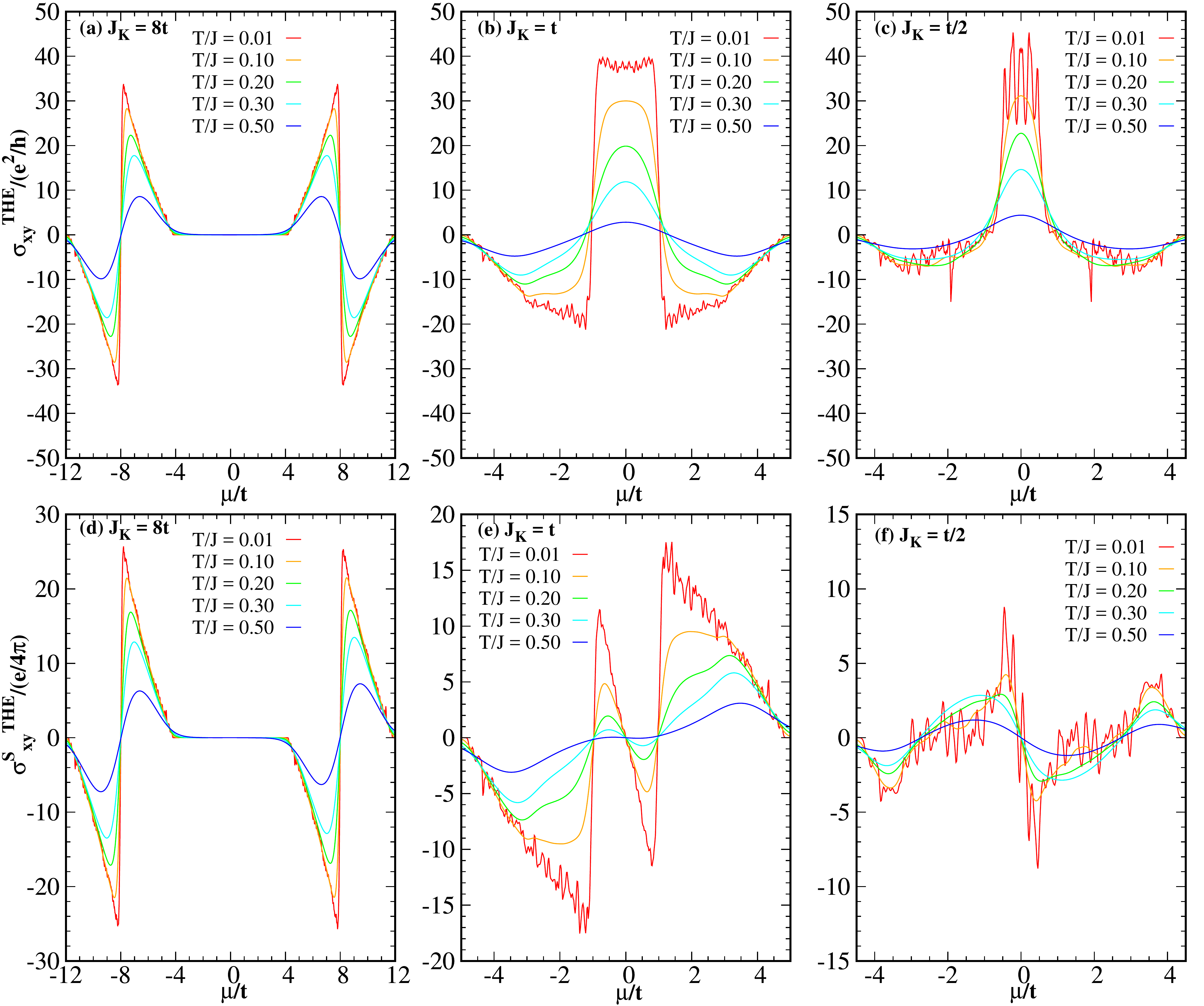}
}
\caption{\label{cond4}
Color online: Behavior of the topological charge Hall conductivity (a-c) and spin Hall conductivity (d-f)
with varying chemical potential and temperature in the SkX phase ($B/J = 0.4$) for indicated $J_K$ values.
}
\end{figure*}
% -------------------------------------------------------------------

With increasing temperature thermal fluctuations destabilize the skyrmion phase further.
As a result, the Hall conductivity reduces continuously and vanishes for $T \approx J$.
The temperature at which the Hall conductivity reduces by $95\%$ 
of the ground state value is identified as the crossover temperature
(see Fig.\ref{cond1}).
In the spiral and the fully polarized phases, Hall conductivity is zero in the ground state.
However, near the phase boundaries of the skyrmion phases, 
thermal fluctuations give rise to small, non-zero values of 
Hall conductivity even in the spiral and the fully polarized phases.
An interesting consequence is a reentrant behavior of the
topological Hall conductivity that is observed
with increasing temperature for a spiral or fully polarized ground state 
close to the boundary with the skyrmion ground state. At low temperatures, 
the Hall conductivity vanishes due to zero net chirality for the ground state
spin configuration. With increasing temperature, thermal fluctuations induce
isolated spin clusters with non-vanishing net chirality which drives a finite THE. 
Finally, at high temperatures, any spin ordering is suppressed by large thermal fluctuations, 
and consequently, the topological contribution to the Hall conductivity vanishes.

We now turn our attention to the change in the behavior of transverse
conductivity as the coupling between the conduction electrons and local
moments is varied from strong to intermediate to weak. The results are
shown in Fig.\ref{cond2}. In the strong coupling regime, the plot consists of
two well-separated blocks corresponding to electrons aligned parallel 
and anti-parallel to the local moments. The two segments exhibit
similar features, but with opposite signs reflecting the opposite spin
alignment of the electrons. The magnitude of the Hall conductivity
increases in steps with increasing Fermi energy, before changing
sharply from a large negative value to a large positive value as the 
Fermi level crosses the van Hove singularity for the lower energy
band corresponding to spin parallel electrons. Further increase in
Fermi energy sees the conductivity decrease in steps of $e^2/h$
to zero at complete band filling for the spin-aligned electrons. 
The sudden change in the nature
of the Hall conductivity is driven by the change in the nature of 
Fermi surface changes from electron-like to hole-like as the chemical 
potential crosses the van Hove singularity.The behavior for the upper band 
follows a similar behavior with opposite signs for the conductivity.
The variation of the Hall conductivity with Fermi energy remains 
qualitatively similar for all $J_K > 4t$ where the two sets of energy 
bands are completely separated. 
For intermediate to weak coupling strengths ($J_K\lesssim t$), 
the two sets of bands start to overlap. 
The electrons are no longer strictly parallel or anti parallel
to the local moments and the energy eigenstates comprise of contributions
from both species. This Hall conductivity remains finite, and follows
the same generic dependence on Fermi energy, but the conductivity plateaus
are suppressed. The sharp change in the sign of $\sigma^\text{THE}_{xy}$
remains unchanged since it arises due to the van Hove singularities
and not the strength of the electron-local spin interaction.

Next we discuss the behavior of the charge and spin Hall conductivities in the skyrmion phase.  
While both the Hall conductivities are interlinked, 
we find interesting differences in their behavior in the skyrmion phase.
Fig.\ref{cond4} shows the variation of the charge and spin Hall conductivities 
with changing chemical potential for different values of $J_K$.
It can be clearly seen that the charge Hall conductivity is symmetric for positive and negative values 
of the chemical potential. However, this is not the case for the spin Hall conductivity. 
In the strong coupling limit ($J_K \gg t$), the energy bands for local spin-aligned and 
anti-aligned electrons are separated by a wide gap and as a result, the charge and spin
Hall conductivities follow one another closely since only one species of electrons
contribute to the Hall conductivities. 
In the ground state, both the charge and spin Hall conductivities show quantized conductivity plateaus, 
exhibit a sharp jump and change in the sign as the chemical potential is tuned across the van Hove singularities.
With increasing temperature we observe the disappearance of the conductivity plateaus and 
a systematic reduction in the conductivity values for any given chemical potential, 
which can be attributed to the thermal fluctuations. However, the change is sign of the conductivities 
at the van Hove singularities remains true.
At intermediate and weak coupling, the
local spin-aligned and anti-aligned bands overlap. The electron spin states hybridize 
and the spins of itinerant electrons no longer follow the local spins, 
i.e., the electrons spins are not simply aligned or anti-aligned to the local moments. 
The energy eigenstates have finite overlap with both electronic spin states. As a consequence, the charge and spin
Hall conductivities are decoupled from each other. 
As observed for $J_K \gg t$, both the ground state, 
charge and spin Hall conductivities exhibit quantized conductivity plateaus 
separated by non-monotonic variation as the chemical potential is tuned, and 
exhibit a sharp jump and change in the sign 
as the chemical potential is tuned across the van Hove singularities.
However, in a sharp contrast to the $J_K \gg t$ regime, for $J_K \leq t$, the charge Hall conductivity is finite at $\mu=0$.
This can be explained due to the existence of large density of states at $\mu=0$ in the $J_K \leq t$ regime.  
The spin Hall conductivity vanishes at $\mu=0$ irrespective of the $J_K/t$ value,
With increasing temperature the conductivity plateaus disappear and the conductivity values 
reduce gradually for a given value of the chemical potential.

\section{Summary}\label{sec:summary}
We have studied the emergence of field induced skyrmions 
in a chiral magnetic system and their effects on electron transport.
Our results show that the interplay of ferromagnetic exchange and DM interaction
in the presence of external magnetic field gives rise to 
several interesting magnetic phases, including the non-collinear
spiral and the non-coplanar skyrmion phases.
The nature of spin textures in these phases and 
the associated phase transitions are analyzed in detail 
with varying magnetic field and temperature. 
In the ground state, the Skyrmion phase consists of a regular
array of Skyrmions forming a Skyrmion crystal. With increasing
temperature, the Skyrmion crystal melts and over a finite
range of temperature, the periodic array of regular Skyrmions is
replaced by a random distribution of distorted spin chiral configurations. 
We term this phase  Skyrmion liquid. Finally at high temperatures,
the Skyrmion liquid melts to a paramagnetic state.
Both the skyrmion crystal and liquid phases are shown to possess 
non-trivial spin-spin correlations and finite spin chirality.
Complementing the magnetic properties, we have studied the band 
structure and the transport properties
of itinerant electrons coupled to different magnetic phases.
The skyrmion phases exhibit a topological Hall effect and 
show finite charge and spin Hall conductivitities, driven by
the coupling of the itinerant electrons to the non-coplanar spin
texture that generates a real space Berry curvature and associated
effective local magnetic field.
We have studied in detail the variation of these quantities with changing temperature,
carrier density and the tuning of the coupling between itinerant electron and localized spin. 
The appearance of conductivity plateaus in the transverse conductivity 
in the skyrmion crystal phase is discussed in detail, 
including an understanding of the unusual behavior 
in the vicinity of van Hove singularities. The effects of 
finite band dispersion on transverse conductivity are also analyzed.
Our results are expected to help in identifying materials exhibiting 
skyrmion phases  
based on magnetic and transport properties complementing the currently used  complex imaging techniques.

\begin{acknowledgments}
We acknowledge use of the HPCC cluster at NTU, Singapore
and the NSCC ASPIRE1 cluster in Singapore for our numerical simulations. 
The work is partially supported by Grant No. MOE2014-T2-2-112 
of the Ministry of Education, Singapore. 
\end{acknowledgments}

\bibliographystyle{apsrev4-1}
\bibliography{skyrmion_ferro}
\end{document}